INFORMAL EDUCATION IS ESSENTIAL TO PHYSICS:

# Findings of the 2024 JNIPER Summit and Recommendations for Action

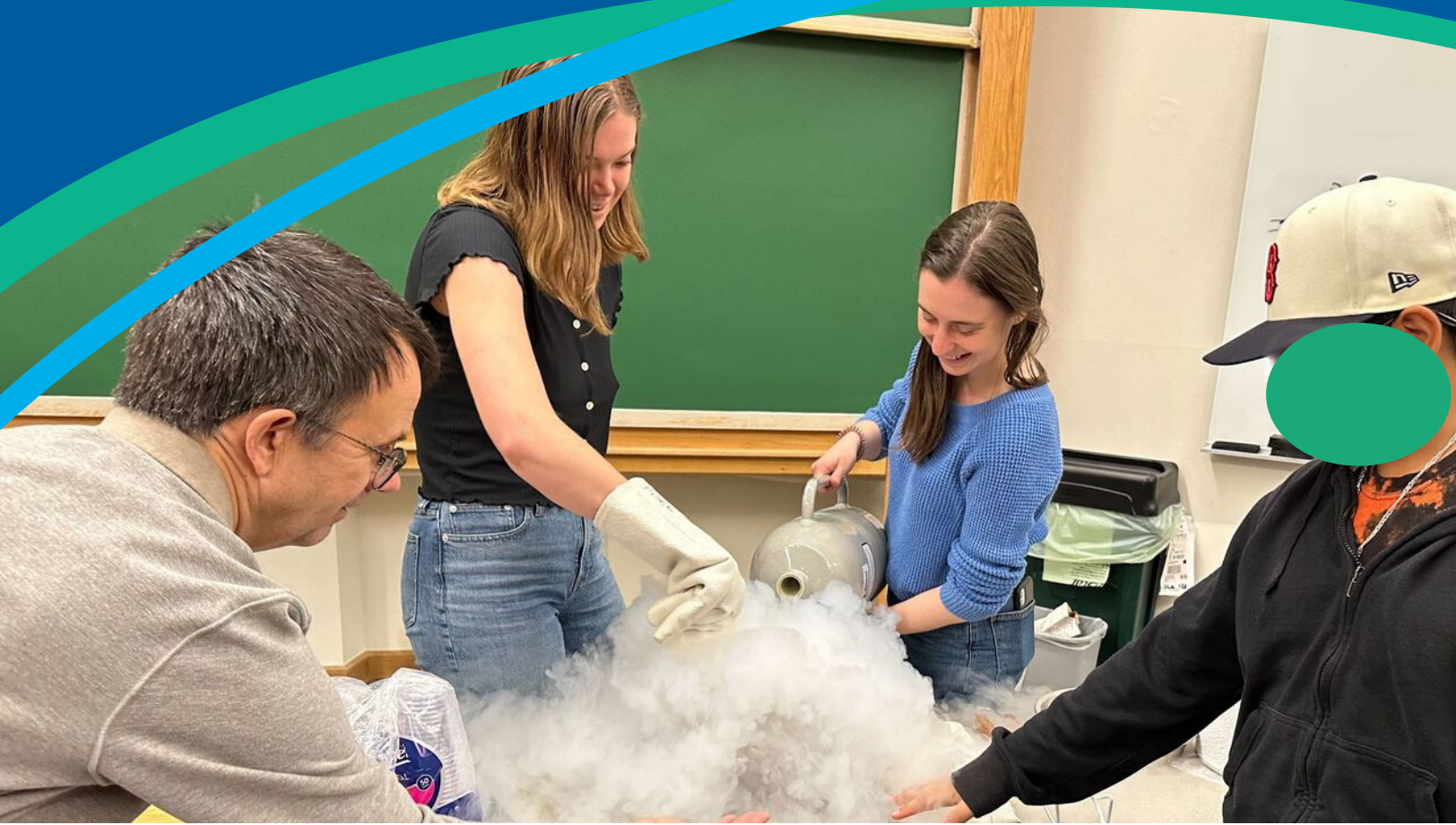

Published July 2025

APS Advancing Physics

# Authors

Alexandra C. Lau[1]\*, Jessica R. Hoehn[2], Michael B. Bennett[2], Claudia Fracchiolla[1], Kathleen Hinko[19], Noah Finkelstein[2], Jacqueline Acres[3], Lindsey D. Anderson[2], Shane D. Bergin[4], Cherie Bornhorst[5], Turhan K. Carroll[6], Michael Gregory[20,21], Cameron Hares[7], E. L. Hazlett[8], Meghan Healy[1], Erik A Herman[9], Lindsay R. House[10], Michele W. McColgan[11], Brad McLain[12], Azar Panah[13], Sarah A. Perdue[14], Jonathan D. Perry[10], Brean E. Prefontaine[15], Nicole Schrode[1], Michael S. Smith[16], Bryan Stanley[17], Shannon Swilley Greco[18], Jen Tuttle Parsons[1], and the broader JNIPER Community

[1]American Physical Society, [2]University of Colorado Boulder, [3]Whitman College, [4]School of Education, University College Dublin, Ireland, [5]Colorado State University, [6]University of Georgia Department of Workforce Education and Instructional Technology, [7]University of Colorado Boulder PISEC, [8]St. Olaf College, [9]Physics Bus, [10]University of Texas at Austin, [11]Siena College, [12]Center for STEM Learning, University of Colorado Boulder, [13]The George Washington University, [14]University of Wisconsin–Madison, Department of Physics, [15]Duke University,[16]Stellar Science Solutions, [17]Lansing Community College, [18]Princeton Plasma Physics Laboratory, [19]Michigan State University, [20]European Physical Society, [21]University of Valencia

\* corresponding author: lau@aps.org

# Author Contributions

All authors contributed to the recommendations included in this white paper. All authors reviewed the paper draft and had the opportunity to provide feedback. ACL, JRH, MBB, CF, KH, and NF conceptualized the JNIPER Summit and organized the discussions around the change levers reported on here. ACL and JRH led project supervision and administration; wrote the original draft; incorporated edits and revisions; and prepared the manuscript for posting publicly. MBB and CF contributed to writing the plenary summaries and the section on Integrating Research-Based Practices in Informal Physics Education. MBB and CF also offered formative feedback at multiple stages in the writing process. NF reviewed early drafts of the paper. TKC contributed to the section on Integrating Research-Based Practices in Informal Physics Education, and ELH and SDB contributed to the opening framing and Section II.

## Acknowledgements

We thank Rachel Larrat, APS Creative Services Manager, for the design and layout of this white paper.

We also thank all JNIPER community members for their participation in an August 2024, virtual coffee hour where initial themes from the Summit were discussed.

## Funding Acknowledgement

The JNIPER Summit, findings from which are reported here, was funded in part by the JILA NSF Physics Frontier Center and the American Physical Society, with additional support from the University of Colorado Office for Research & Innovation and University of Colorado Boulder Center for STEM Learning.

# Executive Summary

In our rapidly evolving world, it is becoming increasingly clear that the wellbeing of both society and science depends on authentic partnership whereby science is practiced in collaboration with, and in service of, the public good. One needs only to look at the public health response to the COVID-19 pandemic, or more recently, the dangerous proliferation of misinformation following Hurricanes Helene and Milton in the United States, to understand the grave need for scientists to build trusting relationships with the public. Effective public engagement with, and in, science is essential for scientists to be trusted and also trustworthy.

Advances in science and technology, from artificial intelligence to mRNA vaccines to fusion energy, will deeply impact our everyday lives and demand input from an informed public. Everyone benefits from having a basic understanding of science principles and methods, and the public deserves to know the findings of taxpayer-funded research. Beyond being informed, the public must have a voice in the scientific enterprise in order for science to meaningfully respond to the various socio-scientific issues we face. Unfortunately, we as scientists are not providing adequate opportunities for society to engage with science, nor are we preparing scientists to authentically engage with society (of which they are part).

This white paper focuses specifically on physicists' engagement with public audiences and the public's engagement with physics. This disciplinary focus aligns with the background and professional identities of many of the paper's authors. Physicists have a complicated history of engaging with the public [1–4], and our modern physics culture often actively disincentivizes public engagement except in service of advertisement or recruitment to physics departments and labs [5–9]. While not everyone needs to become a physicist, everyone should have the access and opportunity to engage with physics. Both physicists and members of the public benefit from engaging with each other. **Therefore, informal physics education (IPE) — a form of public engagement or science outreach — should be an integral and valued scholarly practice within all physics-related organizations.** This is not to say that every physicist has to engage in the same way. IPE includes a range of activities such as science camps, afterschool programs, museum exhibits, demo shows, star-parties, public lectures, and more [10–15]; physicists, as a collective, should engage in the full spectrum of IPE activities, but an individual's level of participation will naturally shift based on their goals, experience, and resources. **In this paper, we explore what it means for IPE to be recognized as an essential practice within physics and what it will take to achieve this culture change in the discipline.**

Most of the learning in one's life, including physics learning, takes place outside of formal schooling [15]. As the primary mode in which publics engage with physics, IPE plays a crucial role in promoting public understanding, trust, and support for science [15]. In responsibly engaging the public through informed, research-based practices, physics earns its social license to operate. For the public, an increased trust in scientific results and scientists can promote action aligned with scientific evidence, and therefore more likely to lead to the expected outcome [16]. IPE also contributes to social progress by fostering curiosity, critical thinking, and a shared knowledge base, enabling local communities[1] to collaboratively address challenges and innovate solutions [15,19]. This disposition toward exploration and learning not only supports individual decision-making, but also cultivates a society that values scientific insights and evidence-based progress. IPE serves as a crucial bridge, allowing scientists to build lasting relationships with the publics they serve and belong to, reinforcing trust and collaboration in advancing the common good.

IPE plays a vital role in expanding access to physics and engaging a broad range of learners from diverse backgrounds. This is important because physics must include diverse perspectives, bringing the breadth of creativity and talent from all members of society, in order to set the vision for the future of physics, and to reach the full potential of science and technology to benefit society [20,21]. IPE can play a crucial role in driving this openness by providing accessible, flexible spaces where participants feel welcomed and valued as they explore their own participation in the field. Indeed, IPE creates spaces where participants can explore and engage with science as their authentic selves, and this helps spark interest in science, supports curiosity-driven inquiry, and promotes physics identity development [22–32]. These benefits also serve physicists who act as facilitators in IPE spaces, and thus, IPE promotes both access to, and retention in, the field [28,29,33,34]. IPE also strengthens "traditional" areas of physics practice, including physicists' teaching and communication skills, and their research [28–31,33,35–39].

**Despite its transformative potential, IPE remains on the fringe of most physics institutions** (universities, national labs, etc.), often led by facilitators in precarious positions, such as students and untenured staff, who rarely hold positions of power [40–43]. Within the physics community, IPE practitioners and researchers receive little recognition or resources for their efforts. **The failure to recognize IPE as an integral part of physics — one that belongs as a central practice in all physics departments, organizations, labs, and national efforts[2] — not only undervalues the critical contributions of IPE, but also limits**

---

[1] We use the phrases "local communities," "society," "public(s)," and "various publics" to refer to people who generally reside outside of the professional science ecosystem. These groups are often referred to as the "general public," but wherever possible we endeavor to use the plural "publics" in recognition that a monolithic "general public" does not exist [17,18]. Rather, various publics have their own localized interests, values, history, and context.

[2] While we call for IPE to be a central practice in all physics institutions, the actual site of the engagement may often be located outside of these institutions, at venues such as planetariums, museums, community centers,

**the potential of the physics discipline.**

The recommendations provided in this paper for achieving a culture change in physics, such that IPE becomes recognized as an essential physics practice, are derived from a Summit hosted by the Joint Network for Informal Physics Education and Research (JNIPER) in June 2024 [44]. JNIPER is an international community of practice [45,46] for facilitators, designers, and scholars of informal physics education activities and programs. At the Summit, we discussed the current status of IPE in physics institutions and we explored various mechanisms for producing the aforementioned culture change. These conversations resulted in a set of recommendations for establishing IPE as a central physics practice. Together, these recommendations provide a roadmap forward for the IPE community and its supporters.

The recommendations are organized around three levers for promoting culture change and four types of change actors.

The three levers for promoting the culture change are structures; engagement of interested, influential, and impacted parties; and integrating research-based practices in IPE (**Figure 1**).

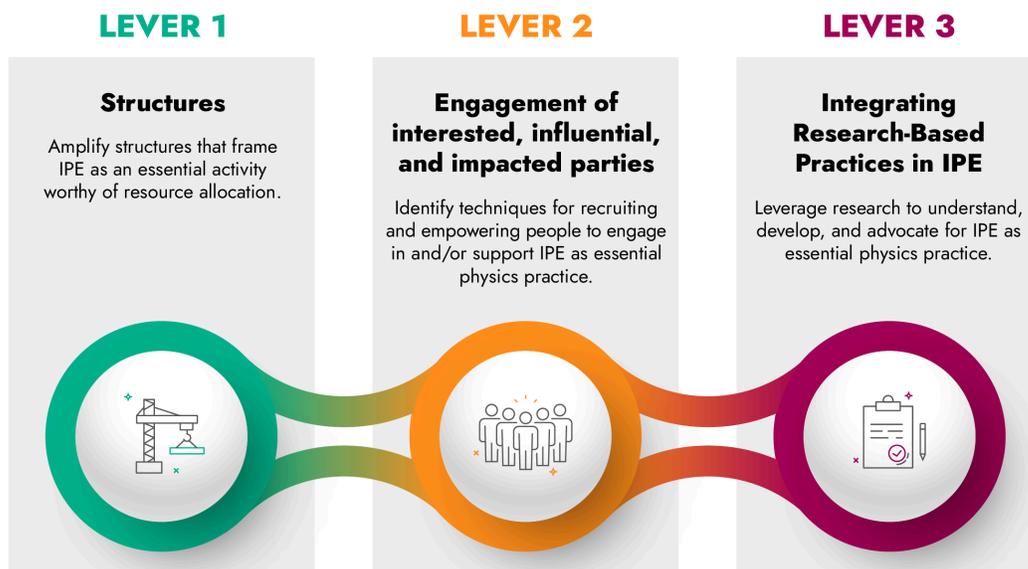

**Figure 1**: Three levers for promoting culture change such that IPE is recognized as an essential physics practice.

libraries, and afterschool programs. Staff at these locations also run IPE activities independent of formal connections to physics institutions. Limited resources is a common issue across all sites for IPE.

Each lever has associated recommendations for action. Culture change requires both top-down and bottom-up effort, and, accordingly, we include recommendations directed to multiple levels of social/institutional power:

- individuals;
- departments and institutions;
- topical groups such as JNIPER; and
- funders and (inter)national organizations.

We encourage the reader to direct their attention especially to the [recommendations at the level(s) where they have the power to act](#). **Our clarion call is for members and supporters of the IPE community to choose their own set of recommendations to prioritize and to set forth a roadmap for implementation.** Together, we can establish IPE as a central physics practice, ultimately leading to a deeper connection between physics and society, strengthening our mutual potential and impact for good.

## A note on audience:

We intend for this paper to be read by IPE practitioners, IPE researchers, IPE organizations/organizers, and IPE funders. People and entities who are supportive of IPE, but not directly involved, may also find this paper of use. We include arguments in support of IPE with the goal of summarizing these in one place for our audience to be able to use in advocating for IPE. When engaging in this advocacy, the messaging should be framed and contextualized in a way that fits the intended audience (see recommendation P.18). The recommendations in this paper include suggestions on how to increase support for IPE, but the recommendations assume an agent already supportive of (or at least not opposed to) IPE. We frame this paper as a call to action and a roadmap for the IPE community and its supporters, given our observations and experiences, supported by the findings of the JNIPER Summit, that IPE lacks a guiding community document to direct change.

We also note that while JNIPER supports a global community, we are a US-based organization with US-based leadership. This means that many of the examples and context we provide in this paper are US-centric. The level of support for IPE depends on the country[3], but the culture change we call for is not just a US issue. JNIPER members from outside the US have provided extensive and valuable perspective not only during the JNIPER Summit but also on the arguments and recommendations for culture change included in this text. We thus argue that overall, this paper's recommendations are broadly applicable beyond the US, but we acknowledge that international readers may need to make adaptations to fit their local context.

---

[3] Ireland, for one, is a positive example of institutionalized support for public engagement [47].

## Table of contents:

In this report, we first define informal physics education and provide background context on JNIPER and the Summit conversations. We build on the key findings from the Summit to motivate culture change within physics and envision what physics would look like if IPE was recognized as an integral practice within the discipline. We identify three levers for promoting the culture change: structures supporting IPE; engagement of interested, influential, and/or impacted parties; and integration of research-based IPE practices. Each lever is accompanied by associated recommendations for action directed at individuals, departments and institutions, topical groups such as JNIPER, and funders and (inter)national organizations.

# I. Introduction

We are living in a civic science era; this means that many of our societal challenges require solutions informed by science, and many of the questions scientists are exploring have social implications, or at the very least, benefit from engaging voices outside of the science community [48–50]. It is insufficient for scientists to "let the science speak for itself," and to communicate only with their science colleagues [16]. Beyond sharing science results with various publics, broader engagement of society throughout the scientific process is needed [49]. This white paper focuses on public engagement with physics, and physicists' engagement with the public; however, we expect that at least some of the included recommendations will resonate with those seeking to elevate public engagement in other STEM disciplines.

In order to reach the full civic and scientific potential of physics, we argue that informal physics education (also referred to as public engagement or outreach) should be recognized as an essential practice within physics. That is, **engaging in informal physics education (IPE) is part of what it means to "do physics."** A culture change in physics is necessary to establish this perspective as the value of IPE is not widely acknowledged or prioritized in the physics community, in its norms, and in the structures in which it operates.

## A. Defining informal physics education

Broadly, informal physics education (IPE) is characterized by physics-related educational activity that takes place outside of formal classrooms and often aims to foster curiosity and excitement in physics while engaging a broad base of participants [10,12–15]. Further, many IPE activities involve mutual engagement of facilitators and participants exploring physics phenomena and topics in ways that are learner-centered and learner-led. These activities range from science camps, afterschool programs, museum exhibits, demo shows, star-parties, videos, podcasts, and more. In this paper, we use the terms "public engagement" or "informal education" rather than "outreach," following ref. [12]. In particular, this usage supports our efforts to highlight the elements of engaging the public and informal learning that are active, bidirectional, and relational. We note, however, that usage is not monolithic and practitioners may opt to use any of the above terms to refer to activity that meets the above criteria. Regardless of the label we (the authors) use for these activities [51], this paper emphasizes an approach that centers mutual engagement and genuine partnership between scientists and local communities and promotes positive interactions between members of the public and the scientists/IPE professionals engaging with them.

Informal physics education takes place in, and is organized by, physics institutions, such as physics departments and national labs, as well as in interdisciplinary entities, such as science centers and libraries. Therefore, facilitators of IPE may consider themselves physicists, or they may more closely identify with another discipline. The recommendations presented in this paper mostly focus on IPE that is facilitated by those who identify as physicists or IPE that is organized/affiliated with physics institutions. This focus is not meant to diminish the existing offices, centers, and people who have promoted science engagement with the public for years; rather, the more narrow physics context provides a critical spotlight on how the discipline can elevate its public engagement work, amplifying existing efforts and identifying unique contribution paths.

## B. Context— What is JNIPER?

The recommendations for establishing IPE as an essential physics practice are derived from a Summit hosted by the Joint Network for Informal Physics Education and Research (JNIPER) in June 2024 [44]. JNIPER is an international community of practice for facilitators, designers, and scholars of informal physics education activities and programs [44]. A community of practice is a group of people who share a common interest in a topic or activity and who are actively working together to deepen their knowledge and expertise in that area by engaging in regular, structured interactions [45,46]. Communities of practice are characterized by: a shared domain or goal that creates purpose and identity; a supportive relationship among the community members in their interest for growth and achieving the goal together; and the development of a set of practices, tools, experiences, and methods that they co-create to achieve their goals. Through collaboration, members of a community of practice enhance their knowledge and skills while collectively advancing best practices and innovation in their field.

JNIPER officially formed in the Fall of 2022, although the conceptualization of an informal physics education research (IPER) community of practice dates back to at least 2017. JNIPER is supported by the American Physical Society (APS) in collaboration with Michigan State University, University of Colorado Boulder, and the JILA Physics Frontier Center. JNIPER seeks to promote a culture where public engagement is an integral practice of physics, fostering a collaborative environment that enhances participation and accessibility, particularly for those historically excluded from physics.

**Table 1**: Elements of the JNIPER community of practice

| **JNIPER Community of Practice** ||
|---|---|
| Domain ("what") | Connection, collaboration, and learning around effective and impactful ways to engage with the public on physics topics, and how to advocate for informal physics education work |
| Community ("who") | IPE practitioners, researchers, and enthusiasts from around the world including students, post-docs, faculty, staff, public engagement |

|  | professionals, etc. from academic and non-academic institutions |
|---|---|
| Practice ("how") | Monthly coffee hours, workshops, conference sessions, professional development program, networking, and online collaboration |

JNIPER promotes connection, collaboration, and learning around effective and impactful ways to engage with the public on physics topics, and how to advocate for informal physics education work (see Table 1). We define effective and impactful IPE as most often taking the form of a partnership approach [52,53]. This partnership perspective prioritizes mutual engagement, where scientists and communities learn from one another, enhancing public understanding of physics while simultaneously providing scientists with insights into public perspectives. In this way, JNIPER functions not only as a support network, but also as a collaborative force driving forward a more inclusive and impactful approach to physics engagement.

JNIPER brings together a wide range of participants, including students, post-docs, faculty, and public engagement professionals, all committed to advancing informal physics education. The JNIPER community welcomes members with a range of public engagement backgrounds, from novice to experienced. Currently, JNIPER has 325 members across over 35 countries, with the majority from the United States. JNIPER members engage with a wide range of publics, from K-12 students and undergraduates, to families, educators, and seniors. The network places a special emphasis on reaching groups who are historically underserved by their connection to the physics discipline. This focus is needed in order to fully incorporate the complete range of intellectual perspectives and experiences along a number of different axes, including, but not limited to, race/ethnicity, gender, and socioeconomic status. The JNIPER community connects with informal physics and public engagement programs across the world, sharing strategies that help members more effectively reach their audiences and achieve their goals.

Through JNIPER, members collaboratively develop tools, resources, and advocacy approaches that encourage physicists to build reciprocal relationships with the public. JNIPER hosts monthly virtual coffee hours featuring the work of our members; these events offer a consistent platform for face-to-face interaction and educational experiences, as well as recognition for our members' work. We also run public engagement workshops, conference sessions, and a science communication professional development program for students. We have a community Slack channel, mailing list, and newsletter to promote connection and bring visibility to the public engagement work of our members.

By advocating for informal physics education as a respected and supported part of a physicist's role, JNIPER promotes practices that elevate both the reach and impact of public engagement efforts.

## C. Summary of JNIPER Summit Conversations

In June 2024, we gathered a subgroup of forty-two JNIPER members from over twenty-five institutions for an in-person Summit to discuss concrete steps for fostering a cultural shift in physics such that public engagement is widely recognized as an essential disciplinary practice. This Summit, which took place at the JILA Physics Frontier Center at University of Colorado Boulder over two days, marked a shift from internal leadership discussion of this topic to broad community engagement on strategy and implementation plans. We designed the event to reflect and strengthen the elements of JNIPER as a community of practice. The gathering sought to build a deliberative forum to enact the following values of:

- **Community building**: Including ample time and space to meet and connect, and engaging in community-led activities to build towards common vision and identity. Crucial to building community, we emphasized social connection and designed activities to recognize and value a common humanity and the span of experiences of all participants (e.g., enjoying meals together, gathering in an inspiring meeting place, valuing humor, prioritizing emergent conversations).
- **Creativity**: Moving beyond scripted presentations or typical production of lists and narrative, to include visual representations, creative arts, multiple perspectives, and multiple discussion formats.
- **Engaging multiple voices**: Participants represented a wide array of institutions, roles, perspectives, and experiences. We provided multiple avenues of engagement and aimed for each individual to contribute substantially to the discussions. Even the plenaries were framed as 'testimonies', intentionally selected to serve as a springboard inviting participants to contextualize, respond to, and build upon from their own perspectives.
- **Scholarship in all its forms**: building on what is known, placing value on both findings from scholarly literature and on-the-ground practical experience, while at the same time recognizing the limitations and biases of all ways of knowing.

Ultimately, the Summit valued and sought emergent conversations, questions, and ideas by building on the collective prior knowledge, wisdom, lived experience, and expertise of the participants.

To these ends, the Summit included a welcome from Eric Cornell (physicist, Nobel Laureate, and champion of IPE within the JILA Physics Frontier Center), plenary sessions, small group discussions, reflective and creative activities, and group meals. Our plenary speakers included Brooke Smith (the Kavli Foundation), speaking on science communication and capacity building; Eve Klein (Association of Science and Technology Centers), presenting research on how the American public thinks about engaging with science; Shannon Greco (Princeton Plasma Physics Laboratory), sharing a case study of an institution with a strong culture of public engagement; and Dr. Katie Hinko (Michigan State University), discussing how research on IPE can be used to develop and advocate for IPE programs. Summaries of the plenary talks are included in Appendix 1. You can view recordings of the talks on the APS YouTube channel [54].

## II. Informal Physics Education as Essential Practice, Integral to the Physics Discipline

In order to support the IPE community in engaging in conversations about the value of IPE, this section explores what it means for IPE to be recognized as an essential physics practice. We enumerate a range of benefits, for physicists and for broader society, that the reader can use as entry points into conversation with various parties about establishing IPE as an essential physics practice. This range of benefits is currently not widely known or embraced by the physics discipline overall, which is why a culture change is needed, as will be discussed in Section III.

Recognizing IPE as an essential practice follows a history of physicists (re)negotiating disciplinary boundaries. The field has defined and redefined what counts as physics, which provides both hope and precedent for the culture change we call for [55]. We have seen how IPE can function as an essential component of scientific organizations. For example, NASA conceptualized public engagement as a key component for garnering the American public's support during the space race and shuttle program [56]. We can redefine physics to include IPE as an essential practice; in turn, IPE, as a space that encourages exploration and creativity, has its own potential to expand the bounds of what counts as physics, and therefore, who can do physics [57].

During the Summit, attendees created visual depictions of what current physics culture looks like and what it would look like if IPE was core to the discipline of physics. Summit attendees emphasized the difficulty of achieving full participation in physics as a discipline, using metaphors like a "treacherous mountain to climb" or "physics as the desert island" (Figure 2). Summit participants explored how IPE can be a bridge to that desert island, allowing people to move on and off the island, or a gentle slope on the mountain, permitting broader access and participation.

Choice about when and how to engage characterize these depictions of physics culture when IPE is incorporated as an essential practice. These characteristics also extend to how the physics community engages with IPE: **engagement will be situationally and contextually dependent**. As noted in the introduction, IPE takes many different forms and there are a myriad of ways to support this work. The unifying feature across this range is recognition and support for the fact that physicists, simply by existing in society as scientists, act as ambassadors for the discipline.

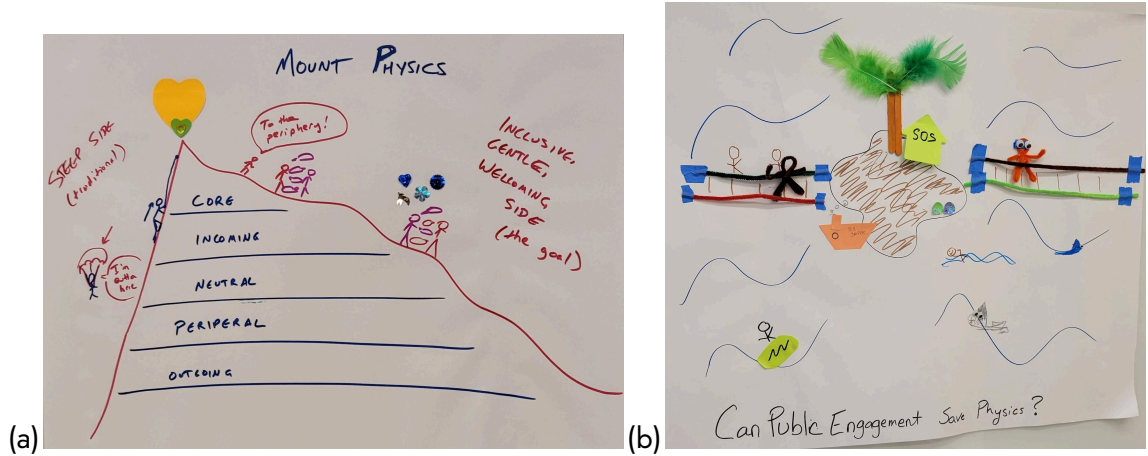

(a) (b)

**Figure 2:** Summit attendees created visual depictions of what current physics culture looks like and what it would look like if IPE was core to the discipline of physics. Diagram (a) depicts "Mount Physics," with the steep side of the mountain representing traditional physics culture. People on this side of the mountain have to crawl their way to the top, or they may decide to parachute off the mountain. The other side of the mountain has a gradual slope and represents the participants' view of what physics culture could be if IPE is recognized as essential. This side of the mountain is more welcoming and open. People can climb up together, and they can also choose how high up the mountain they want to go. Diagram (b) depicts physics as an isolated island. IPE can act as a bridge or lifeboat to the island, helping people easily move on and off, eliminating the isolation.

After exploring what physics would look like with IPE as an essential practice, summit attendees inquired, "Can public engagement save physics [from all the problems with its current culture]?" IPE benefits physics funding, workforce development, science-informed policy, participation, and more. We unpack these elements below. See, too, [Appendix 2](#) for a discussion of calls to action that mirror our call for valuing IPE as an essential physics practice.

### A. IPE is essential physics practice because it helps ensure disciplinary relevance and social license to operate.

As a scholarly community, physicists continuously attend to practices that, in their undertaking, make us physicists — our ways of working; of knowing; of admitting new members. Ensuring these practices are open and fair is essential to the success of the physics community. Care should be given to ensure that boundaries between the community of physicists and those outside the community are porous. There are two reasons for this. First, hard boundaries can make participation exclusive, narrow, and inward looking. Second, hard boundaries can isolate the work of physicists from the world in which it operates. **What is the point of our collective scholarly enterprise if we are just talking to ourselves?**

Informal physics education is key to creating porous boundaries. The messy, liminal, informal spaces on the edges of academic disciplines are how most people engage with those disciplines and, therefore, are essential to the disciplines having meaning and impact. These spaces offer people ways to 'get a feel' for physics — which itself is an inherently messy field of inquiry — without suggesting they formally join the community of physicists, unless they wish to. Informal physics spaces provide people opportunities to know what physics offers, to establish the trustworthiness of claims made by those with physics expertise, and to imagine new ways of looking at things and themselves. Similarly, these spaces offer physicists the opportunity to explore new ways of understanding physical phenomena and their broader relevance and impact, sparked by the novel questions and perspectives that people outside the physics community bring. A community of physicists that does not see informal physics education as essential to their existence risks isolation and irrelevance, and is not equipped to engage in the self-analysis required of authentic science. Spending time at our community boundaries puts us in dialogue with others about the limitations and opportunities of physics knowledge and practices. These conversations result in a clearer sense of purpose for our work and understanding of what we can offer the world.

## B. IPE is an essential physics practice because it fosters trust and supports a society where everyone benefits from scientific and technological advances.

Through IPE, the public gains the opportunity to experience the process of science and to establish the trustworthiness of scientists. According to the most recent survey on trust in scientists from the Pew Research Center, 76% percent of Americans trust scientists to act in the best interest of the public [58]. However, the level of trust is ten points lower than it was pre-COVID19 pandemic and there are deep partisan divides on this topic (66% of Republicans and 88% of Democrats report trust in scientists). Globally, a study across 68 countries finds that there is moderately high trust in scientists (M=3.62/5, SD= 0.70), but there is still much room for improvement [59]. For example, when examining the four constructs of trust (competence, integrity, benevolence, and openness) in the global data set, the study finds that perceptions of scientists' competence is high, but perceptions of their integrity, benevolence, and openness are moderate.

These findings are concerning given the growing difficulty of distinguishing reliable information in today's digital media ecosystem. Mis- and disinformation abound on health and science topics, and at the same time, advances in STEM technology such as artificial intelligence have made it easier to create and spread disinformation. The World Economic Forum lists misinformation and disinformation as the greatest global threat over the next two years, and places it in the top five threats over the next ten years [60].

A Research!America survey from 2025 indicates one possible explanation for the lack of trust: in their national sample, only 32.5% of respondents said they could name a living scientist [61]. Given that there are at least 6.6 million scientists working in the US [62], we argue that scientists are not doing enough to be known and valued by the non-scientist population they serve and are served by. In fact, according to the Pew study, only 45% of U.S. adults think that research scientists are good communicators, and 47% report a sense of elitism among research scientists [63]. Engaging in (effective) IPE is one way for physicists to bridge these divides, learning about the concerns and values of publics, while publics in turn learn about the scientist. This will help foster trust and make it more likely for publics to turn to scientists for information on science and technology matters [16,64].

Encouragingly, the public sees value in science and over 91.5% of respondents in the Research!America study report that it is important for scientists to inform elected officials about their research and its societal impact [65]; 84.6% of respondents said it should be part of a scientist's job to inform the public about these topics [66]. In other words, the majority of the public thinks IPE should be a core part of what physicists do. From an economic accountability standpoint, it makes sense that taxpayers are interested in what their taxes are funding. These statistics further indicate that the public recognizes how science can inform both social and economic policy.

Engaging with the public through partnership-focused IPE experiences can additionally promote science advances that serve local communities, and this in turn will build trust. While many of the challenges facing society today require science and technology to solve, the values and priorities of local communities matter for solutions to be effective. We cannot expect solutions to societal challenges (e.g., safe and sustainable energy supply) to work for broad audiences if only a small subset of the population is consulted in developing them. Science does not operate in a vacuum and it matters who is involved in the process of scientific inquiry. IPE experiences provide a mechanism for such collaboration.

## C. IPE is essential physics practice because it is a gateway for entering into the discipline, and for staying once there.

Currently, physics remains an exclusive community. For example, between 2017-2021, an average of 15% of physics bachelor's degrees in the US were awarded to students identifying as Black and African American, Hispanic and Latino American, American Indian, Alaska Native, Native Hawaiian, or other Pacific Islander [67]. In comparison, these groups comprise 37% of the college-age population in the US [67]. In the same time period, an average of 23% of physics bachelor's degrees in the US were awarded to students identifying as women [68], but the college enrollment rate for women aged 18 - 24 in the US is 44% [69]. This lack of representation is not only a US issue; it pervades physics in many other countries such as Germany [70], Brazil [71,72], India [73], and the UK, including in new and emerging physics fields such as quantum information science and technology [74].

IPE is crucial for sparking interest in science and for supporting learners' development of a physics or science identity [15,22–26,75]. This holds true both for the facilitators and participants in IPE activities [27–31]. IPE has been shown to support recruitment and retention in the field because of its role in sparking interest, connecting with a wider audience, creating a welcoming environment, and fostering science identity. IPE helps people stay in physics because it can both act as a space where students can engage with the discipline as their whole selves [76], while also transforming the physics culture into a more welcoming community. For example, to improve physics department climate, it is recommended to provide IPE experiences for students in the department [34]. IPE is particularly important for recruitment and retention of people from a range of backgrounds, including those less represented in physics [15,22–31,77]. For example, a recent study of the Black and African American physics bachelor's degree recipients from the class of 2022 indicates that their most frequent influences for pursuing a physics degree were out-of-school, informal STEM education experiences [32].

## D. IPE is essential physics practice because it improves physicists' skills and research.

For physicists themselves, facilitating IPE activities can have a positive effect on their teaching skills, professional development, discipline-based identity, sense of belonging within physics, and thus persistence through a physics degree or career [28–31,33,35–39]. For example, one study investigated the effects of participation in informal physics programs on undergraduate and graduate student volunteers and found that students reported positive physics identity development, increased sense of belonging to the physics community, and development of skills in communication, teamwork, networking and design skills [30]. Other studies [28,31,33,39], explore IPE environments as spaces of physics identity development and exploration particularly for IPE facilitators (often students), which is thought to be supportive of persistence in the field. IPE activities can also benefit scientists' research by spurring new ideas and through participatory science initiatives (e.g., community science, also known as citizen science). For example, JNIPER member Christina Love has shared how a community science initiative, "Name that Neutrino," hosted on Zooniverse has helped the IceCube Neutrino Observatory classify neutrino signals [78]. JNIPER member Lindsay House is leading another community science initiative hosted on Zooniverse, called "Dark Energy Explorers" in which participants help classify data from the Hobby-Eberly Telescope Dark Energy Experiment [79,80]. Engaging the public with research can spark novel ideas for physicists to pursue and it can strengthen their own understanding of their research area.

Given all of these benefits, IPE should be recognized as an essential physics practice; it opens the discipline to new people and ideas, strengthens public support and understanding of the field, supports learning and development of physicists and physicists-in-training, and reinforces the discipline's reason for being. **A corollary to this assertion is to consider: can physics survive without IPE?**

## III. Driving Culture Change

Establishing IPE as an essential practice within physics requires a culture change in the discipline. This is because the points enumerated in the previous section are not widely acknowledged or prioritized in the physics community, in its norms, and in the structures in which it operates.

Longstanding research on institutional and organizational change, particularly in higher education, STEM fields, and physics programs, shows that change within physics programs is both a complex and integrated system, distinct from how businesses change [81–88]. Successful change efforts must attend to structures, power, and people. At most institutions, the department serves as the unit of change. Effective transformation requires a grassroots approach that meets a top-down support structure, while integrating evidence-based practices [5,89] Taking an evidence-based approach is not only a good idea, but necessary in order to ensure productive outcomes, to be recognized as a scholarly practice/approach (i.e., to align with the culture of physics practice), and to allow for coordinated action and evaluation across isolated programs.

To drive meaningful change in public engagement within physics, we focus on three critical, interrelated levers: a) structural support, b) engagement of key parties at all levels, and c) implementation of research-based practices (**Figure 3**). Thus, the following analyses and recommendations are directed towards individuals, departments, informal STEM topical groups (e.g., JNIPER), and to (inter)national organizations and infrastructure.

Sustainable changes require collaboration across all levels. Individuals conduct IPE work and departments and science organizations are key spaces where IPE occurs and where norms and values of IPE are established. Meanwhile, IPE efforts ought to be aligned with and supported by broader institutional missions. Finally, informal STEM topical groups (e.g., JNIPER) and (inter)national organizations and infrastructure such as professional societies and funding agencies are necessary to support, coordinate, connect, and advance individualized efforts into a whole that is greater than the sum of the parts.

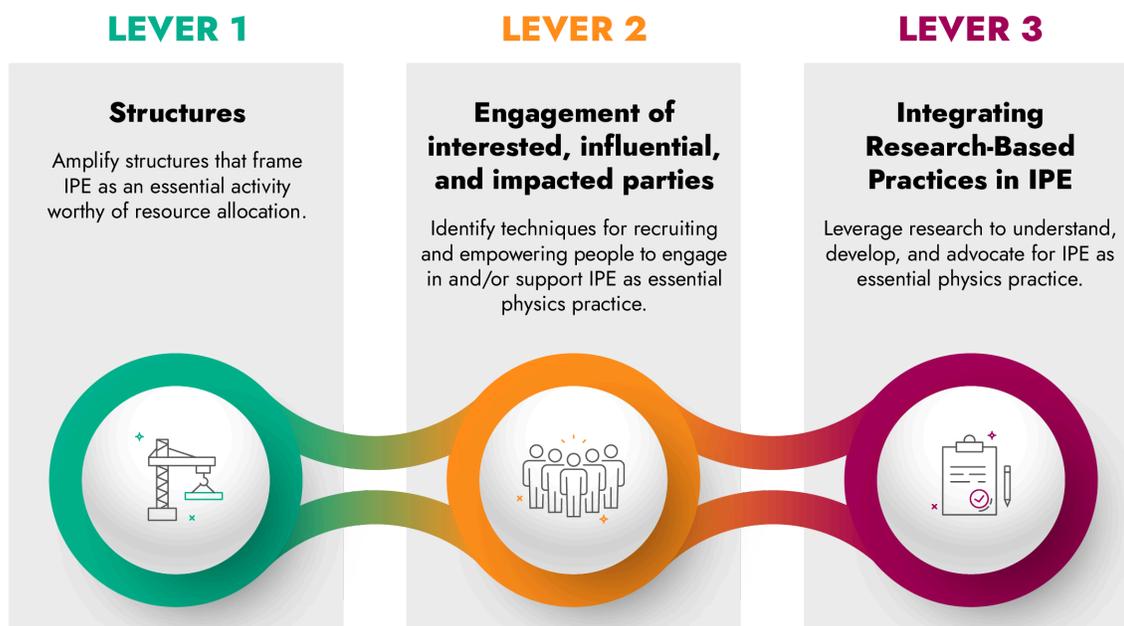

**Figure 3:** Three change levers through which to achieve the desired culture shift whereby IPE is widely recognized as an integral part of the physics discipline.

## A. Change Levers

In the three sections that follow, we describe the "lever" (i.e., the category of activity within the physics discipline in which change may occur) and then report recommendations for the IPE community related to that change lever[4]. The recommendations are organized at four different scales of collective action: a) recommendations for individuals/individual programs, b) recommendations for departments and institutions, c) recommendations for informal STEM topical groups (e.g., JNIPER), and d) recommendations for national and international organizations such as funders, disciplinary societies, or publishers. (See Appendix 4 for the recommendations organized by scales of action, e.g. recommendations for individuals across the three levers, etc.)

---

[4] Some recommendations appear under multiple scales and levers, as applicable. The recommendations are numbered for easier reference, but the order the recommendations are listed in is arbitrary. If a recommendation appears under multiple change agent scales within one lever, it is numbered identically, after the first time it appears; for recommendations that appear across levers, the recommendation is numbered under each lever, and a note indicates the overlap, e.g., "S.19 (*same as R.1)."

**Lever 1: Structures**

Enacting culture change requires attending to the institutional and organizational structures that support and reinforce disciplinary practice. Structures include things like policies, resources, funding, and the organization of how power is distributed [81,84,86]. These are the tangible and intangible factors that can reinforce, or act as barriers to, both individual and group action.

At the Summit, and in subsequent JNIPER community-wide discussions, we asked: What are existing structures at our institutions or across the physics discipline that currently serve to support and recognize IPE? What existing structures currently act as barriers to IPE work? What structures are needed to better position IPE as essential physics practice? Brooke Smith's plenary provided two insights into existing structures, or lack thereof, that impact IPE: (a) a lack of training, and (b) hiring and promotion practices that fail to reward, and in many cases actively discourage, IPE. Summit attendees further identified structures needed to promote scholarship, community-building, and education in support of IPE.

The recommendations are labeled with the convention "S.#," short for "**S**tructures."

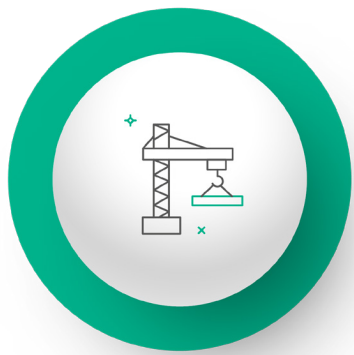

# LEVER 1
# Structures

Amplify structures that frame IPE as an essential activity worthy of resource allocation.

*The recommendations are labeled with the convention "S.#," short for "Structures."*

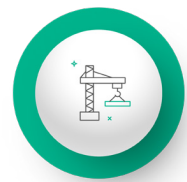

## Recommendations for individuals/individual programs

- **S.1:** Structural changes are often brought about by individual 'champions' who lead organizations in adopting the change. While **we encourage individuals to advocate for the following structural changes**, especially those individuals in positions of relative organizational power, **we also recognize that no one individual ought to be responsible for ensuring the following recommendations come to fruition** within their, or any, organization.

## Recommendations for departments & institutions

- **S.2: Leverage the connection between IPE and service learning** — Some undergraduate education courses are designated as "service learning" courses, in which students must engage in real-world teaching/education experience as part of the course. Physics departments and individual informal physics/STEM programs should partner with these courses to provide opportunities for students to engage in IPE and to simultaneously support the success and sustainability of the informal programs. It would be ideal if this was a structure embedded into departments and undergraduate (or even graduate) programs.

  **Successful examples:** Mobile Making at California State University [63,64] & Eric Hazlett's Analytical Physics 3 course at St Olaf College, which includes an active civic engagement component where students create hands-on demonstrations that are shared at local community events and with students in the St. Olaf TRiO Educational Talent program.

- **S.3: Integrate IPE efforts with other existing efforts to broaden participation** — Integrating IPE with existing initiatives aimed at expanding access and engagement can help departments and institutions make efficient use of established structures rather than creating new ones from scratch. This approach strengthens IPE while also fostering a more welcoming and supportive environment for a wider range of participants in physics. As with all recommendations, implementation will be highly context dependent and should attend to the organizational climate in which the department finds itself.

- **S.4 (*same as P.9): Hiring and promotion policies should include and reward IPE work** — IPE activities should not only be acknowledged in tenure and promotion cases [65], but they should be rewarded and encouraged, contrary to the existing norm in which IPE activities can hinder an individual's opportunities for tenure and promotion. For more context on this recommendation, please see the white paper published by the American Physical Society Committee on Informing the Public [66,67].

  **Successful examples:** At Lansing Community College, faculty contracts include a specified number of hours that must be dedicated on non-teaching assignments, and community outreach and events explicitly count towards this time. Dr. Bryan Stanley, LCC physics faculty, shares that in their hiring interview, they were asked about their community engagement work and plans for future community engagement they wanted to do in the position they were interviewing for.

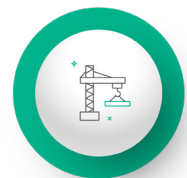

Similarly, the University of Texas at Austin's policy for promotion for professional track faculty (including teaching faculty) includes a statement on your primary area plus another statement on "Contributions to the Academic Enterprise" which is a broad category that can include any substantive additional work, including in IPE [68].

In Ireland, the University College Dublin 'Framework for Faculty,' used by academics in their applications for promotion, includes a specific public engagement dimension. Within this, expectations are set out for faculty with the highest levels associated with public engagement scholarship at international scale [69].

- **S.5: For faculty at academic institutions or other roles with "service" requirements, IPE activities should fulfill said service requirements —** Departments can also consider how IPE could fulfill teaching requirements if, for example, a faculty member includes an IPE component to their physics course (see recommendation S.2).

- **S.6: Provide funding for IPE work at multiple scales —** Funding need not only come from large-scale, national foundations, but should also come locally from departments, universities, etc. Funding mechanisms should prioritize IPE activities that are utilizing evidence-based practices and are designed to effectively engage the audiences of focus. Opportunities for renewable funding will help make IPE programs sustainable, and funding available to students and other junior members of institutions will help support a large population of IPE practitioners. Departments and institutions are well-suited to provide commonly needed (and relatively cheap) resources like materials, room rentals, parking waivers, and stipends for student interns.

- **S.7 (*same as R.14): Establish recognition mechanisms for exemplary IPE programs and practitioners, including students —** Such recognition could include site or practitioner awards, digital badges or professional certificates for completing IPE training, and features in institutional communications. Recognition can be implemented at the department, community of practice (e.g., JNIPER) and international organization (e.g., APS) scale.

- **S.8: Create or expand community engagement/campus extension offices —** These offices are often staffed by experts in community partnerships whose job it is to build relationships with local community members and groups. These offices can be a resource for physics faculty and students seeking to engage in IPE, as well as local audiences interested in STEM. This recommendation relates to institutions ensuring these offices have the capacity and expertise to support these kinds of connections that would be beneficial for physics/STEM public engagement.

- **S.9 (*same as P.12): Obtain the Community Engagement Elective Carnegie Classification for your institution [70]—** This classification requires a campus-wide commitment to partnership with the local community. Applications require detailed examples of academic-community partnerships, such as an IPE program. A push from institutional leaders to obtain this classification for the institution will promote buy-in from multiple interested, influential, and/or impacted parties in IPE.

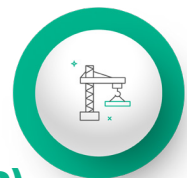

## Recommendations for informal STEM topical groups (e.g., JNIPER)

- **S.10: Host a member directory or other mechanism for facilitating partnerships among IPE practitioners, IPE researchers, and other physicists**

- **S.11: Publish a regular newsletter to support community building, help disseminate members' work, and provide visibility and recognition**

- **S.12 (*same as P.16): Provide toolkits and training to support implementation of partnership-focused IPE programming —** Resources will lower the barrier to starting and improving an IPE initiative because interested parties will not have to "reinvent the wheel." Examples include: a playbook of how IPE practitioners should connect to communities, defining best practices for building partnerships; "cheat sheets" that synthesize research findings and provide suggestions for how IPE practitioners can apply the findings in their work; a list of models or examples for how departments might incorporate IPE in their curriculum, major, or departmental activities and culture.

- **S.13 (*same as P.13): Provide workshops on effective grant writing, impact reporting, and evaluation strategies —** This training can help individuals prepare an IPE-focused grant proposal, and/or help individuals to directly connect their physics research meaningfully to their broader impacts components of their physics research grants.

- **S.14: Set a recognized standard for ethical and effective IPE practice and research —** These standards can be emulated and boosted by science societies, funders, and other IPE partners. Examples of standards include paying and/or recognizing students facilitating IPE programs; cultivating long-term community partnerships; engaging in research-practice partnerships; and establishing clear metrics and evaluation plans. For more on standards for research, see Section III.D.

- **S.7 (*same as R.14): Establish recognition mechanisms for exemplary IPE programs and practitioners, including students —** Such recognition could include site or practitioner awards, digital badges or professional certificates for completing IPE training, and features in institutional communications. Recognition can be implemented at the department, community of practice (e.g., JNIPER) and international organization (e.g., APS) scale.

- **S.15 (*same as R.15): Publish IPE research "digests" —** To help the IPE community and partners stay informed of recent results and findings. This digest should be shared with science societies (e.g., APS, AAPT, AIP) and others outside the immediate IPE community.

- **S.16: Create and disseminate mentorship opportunities —** To mentor people new to IPE into career pathways that incorporate IPE.

    **Successful example:** The JNIPER Fellows program trains a small cohort of students in science communication skills [24]. The students then apply their skills to produce content for APS Public Engagement programs. The cohort design and connection to the broader JNIPER community provide exposure to multiple pathways that incorporate IPE.

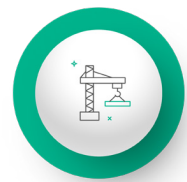

## Recommendations for (inter)national organizations

- **S.17 (*same as R.20): Academic and commercial publishers should establish common publication venues for IPE work —** Existing venues such as *Physical Review PER* [71], the *Journal of STEM Outreach* [72], *Citizen Science: Theory & Practice* [73], *The Physics Teacher* [74], and *Connected Science Learning* [75] should better connect with the IPE community to make them aware of their publication options. Journals should also establish multiple submission types to allow for both research and programmatic articles rooted in experience, practice, and evaluation. Whenever possible, these publication venues should also be made open access.

- **S.7 (*same as R.14): Establish recognition mechanisms for exemplary IPE programs and practitioners, including students** — Such recognition could include site or practitioner awards, digital badges or professional certificates for completing IPE training, and features in institutional communications. Recognition can be implemented at the department, community of practice (e.g., JNIPER) or international organization (e.g., APS) scale.

- **S.18 (*same as R.22): Create dedicated IPE sessions at physics conferences and schedule them in prime slots** — Science societies that run physics conferences can include IPE in their abstract sorting categories, consider IPE practitioners for plenaries, and more.

- **S.19: Funding agencies should support IPE through the following mechanisms —** Require all grants to include a public engagement plan, and providing detailed guidance on evidence-based best practices; Attend to public engagement action and implementation in annual grant reports and reviews, ensuring that quality IPE is valued and is not relegated to a "box-checking" exercise; Fund IPE programs directly; Fund studies of public engagement and science communication trainings to expand knowledge of effective training practices; Direct graduate fellowship awards to include funding for public engagement training and implementation. (This last example is aligned with the call from the Research!America Public Engagement Working Group [76] .)  This recommendation overlaps with R.23.

- **S.20: Create federal structures that support national agencies in growing their participatory public engagement and science communication efforts—** See, for example, the August 2023 Letter from the Presidential Council of Advisors on Science and Technology [77].

## Lever 2: Engagement of interested, influential, and impacted parties

In addition to top-down structures, culture change requires grass-roots efforts and advocacy at each level of the overall physics and IPE ecosystem. Garnering support at every level of power makes for more sustainable culture change because there is no single point of failure. This grass-roots action and advocacy is supported by messaging articulating why IPE is an essential practice of physics in ways that are compelling for different audiences. Summit participants discussed a possible shift in our mindset, however, from one of making a "pitch" about the value of IPE, to one where we are inviting people into *conversation* about its value and benefits.

Within IPE, there are many parties with varying levels of interest, influence, and impact; it is essential to engage each group in the desired culture change. By "interested, influential, and/or impacted parties[5]", we refer to individuals or groups that have an interest in (such as potential gains or losses), can exert influence, and/or are affected by IPE [105]. These parties may or may not be directly involved in IPE. Some of these groups reside outside of traditional physics communities, while others within the physics communities find their IPE work valued in communities outside of physics. Examples of such parties include faculty, students, IPE practitioners, department and university leadership, local city/town members, IPE participants, K-12 teachers, funders, policy makers, and professional societies. These groups often hold different levels of (situationally-dependent) power, and there is variability even within a group (e.g., not all "interested parties" are the same).

Summit participants considered what it would mean for different parties to be "bought-in" to IPE as an essential physics practice. They discussed strategies for achieving that buy-in and empowering people to engage in IPE as an essential physics practice. For example, they identified the need to understand the goals of each party so we can articulate a relevant value proposition. At the same time, it is important to recognize where the goals and interests of these groups overlap because this points to messaging elements that will resonate with multiple parties. For a given group, the benefits of supporting IPE as an essential physics practice must outweigh the possible costs in order for there to be a compelling value proposition.

When achieving buy-in from interested, influential, and/or impacted parties, the messenger matters as much as the message [87,88,107,108]. Shannon Greco's plenary demonstrated the importance of champions for institutionalizing IPE work at Princeton Plasma Physics Lab ([see Appendix 1](#)). Champions can hold various levels of power in the physics ecosystem; being a familiar, trusted member of a group where one is trying to promote a culture change matters more than the official title one holds [16,84,109]. Of course, different groups have different levels of power for change, and those with more power should do more of the lifting.

---

[5] We use the phrase "interested, influential, and/or impacted parties" instead of the more common term "stakeholder" because of the latter term's potentially harmful connotations. Stakeholder is a vague term that can refer to a wide variety of groups who actually have very different interests and levels of power in a situation [105]. For indigenous groups, the term also has roots in colonial practices which removed their rights to land and reduced the relationship between people and land to a purely economic one [105,106].
We thank the attendees who raised concerns (and confusion) about the term "stakeholder" during the Summit. We have now adopted "interested, influential, and/or impacted parties" as a more accurate description of the groups to which we refer.

The recommendations that follow speak to how individuals, departments, informal STEM topical groups, and (inter)national organizations, respectively, can promote buy-in of IPE as an essential physics practice and engagement from various interested, influential, and/or impacted parties. The recommendations are labeled with the convention "P.#," short for "interested, influential, and impacted **P**arties."

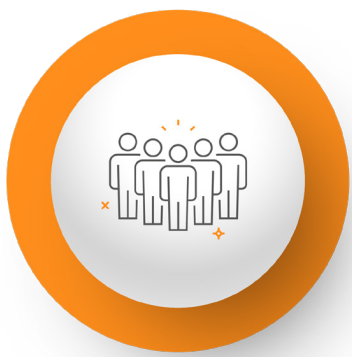

## LEVER 2
# Engagement of interested, influential, and impacted parties

Identify techniques for recruiting and empowering people to engage in and/or support IPE as essential physics practice.

*The recommendations are labeled with the convention "P.#," short for "interested, influential, and impacted Parties."*

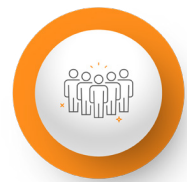

## Recommendations for individuals/individual programs

- **P.1: Track program effectiveness —** University administration, funders, and others often request data on IPE program impact in order to support the effort. Sharing these impacts with interested, influential, and impacted parties can help make the case for the value of IPE. These data can be a mix of quantitative and qualitative metrics. See section III.D on research-based practices for a discussion of a variety of possible metrics.

- **P.2: Align goals with funders' missions —** Funders want to see how the project aligns with their goals. It is important to have a clear plan for the direction you want your project to go, and how support will help you get there. Additionally, proposals should incorporate rich and robust evaluation frameworks and plans for sharing evidence of impact.

- **P.3: Meet with policy makers to advocate for supportive policies and funding for IPE —** Highlight the benefits of IPE for constituencies and their connection to existing educational goals. This can be done via various mechanisms such as policy briefs, roundtables, and 1:1 meetings.

- **P.4: Invite policy makers, administrators, leaders, and faculty to interact with IPE programs —** Observing, showing up, and even participating in IPE programs will increase awareness of programs, their funding needs, and common values. These people can then be recruited as program champions, to bear witness to their importance.

- **P.5: Consistently engage community members at every stage and level of an IPE program —** Community members can be both advocates and avid participants for IPE programs. Co-creating programming with the community and your participants will lead to greater participation and impact. Co-creation can also mean leaning into the "messiness" of multiple voices and priorities.

- **P.6: Integrate IPE into your research goals —** This can take the form of a research project studying the design and impact of an IPE program, or can be incorporated into an area of traditional physics research. For example, you can engage various publics in your data collection and analysis through "community science" platforms. You can also host roundtables with community leaders to understand their needs and how they can shape your research direction.

- **P.7: When engaging in dialogue with different** interested, influential, and/or impacted parties, **articulate a relevant value proposition for supporting IPE —** This is a more general statement of a number of the other recommendations. Find the message that resonates with your audience by understanding their goals and how they connect to those of your program. For example, faculty and department leaders may be interested in the benefits to university students who facilitate IPE programs, while K-12 schools or community organizations may be interested in the benefits for younger students and their families.

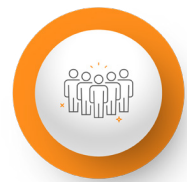

## Recommendations for departments & institutions

- **P.8: Invest in infrastructure that supports public engagement and evaluation —** Reducing administrative and resource barriers facilitates buy-in from faculty, staff, and students. Examples of infrastructure include dedicated space and materials that can be used for IPE programming, partnerships with social scientists for program evaluation, and teaching assistant lines dedicated for facilitating IPE programs.

- **P.9 (*same as S.4): Recognize IPE activities in tenure and promotion processes —** This is important for faculty and staff to buy-in to IPE as a practice they should engage in [65–67].

- **P.10: Integrate IPE activities into the curriculum and provide academic credit for participation in IPE efforts —** This provides an important marker of disciplinary value to students. It is easier for students to buy-in to IPE as an essential disciplinary practice if they get academic credit for their participation. IPE can be incorporated into curricula as part of a service-learning credit, a lab course, a pedagogical/communication requirement, or an independent study.

  **Successful example:** Eric Hazlett's Analytical Physics 3 course at St Olaf College, which includes an active civic engagement component where students create hands-on demonstrations that are shared at local community events and with students in the St. Olaf TRiO Educational Talent program.

- **P.3: Meet with policy makers to advocate for supportive policies and funding for IPE —** Highlight the benefits of IPE for constituencies and their connection to existing educational goals. This can be done via various mechanisms such as policy briefs, roundtables, and 1:1 meetings.

- **P.4: Invite policy makers, administrators, leaders, and faculty to interact with IPE programs —** Observing, showing up, and even participating in IPE programs will increase awareness of programs, their funding needs, and common values. These people can then be recruited as program champions, to bear witness to their importance.

- **P.11: Acknowledge the IPE career pathways that students can take and legitimize the rhetoric around those IPE careers —** Career panels and talks should include examples of physics majors who have gone on to a career in IPE, including the option to pursue the academic track with a research-focus on IPE. Job boards and career resources should also illustrate the many forms an IPE career can take.

- **P.12 (*same as S.9): Obtain the Community Engagement Elective Carnegie Classification for your institution [70] —** This classification requires a campus-wide commitment to partnership with the local community. Applications require detailed examples of academic-community partnerships, such as an IPE program. A push from institutional leaders to obtain this classification for the institution will promote buy-in from multiple interested, influential, and/or impacted parties in IPE.

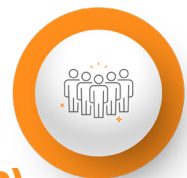

## Recommendations for informal STEM topical groups (e.g., JNIPER)

- **P.13 (*same as S.13): Provide workshops on effective grant writing, impact reporting, and evaluation strategies** — This training can help individuals prepare an IPE-focused grant proposal, and/or help individuals to directly connect their physics research meaningfully to their broader impacts components of their physics research grants. Training in evaluation helps practitioners know what to assess (e.g., participant value, motivations) and helps with messaging to funders and leaders because they often require metrics of IPE program impact.

- **P.14: Highlight successful IPE programs —** Concrete examples of programs that are meeting their goals and those of their community partners help promote buy-in from parties who have not experienced an IPE initiative. Similarly, sharing stories of challenges and how programs have overcome them will add to buy-in.

- **P.15: Highlight institutional practices that uplift IPE work —** Collect concrete examples of institutions that are recognizing the IPE work of their students/staff/faculty and examples how they are supporting the work. This helps facilitate change at similar type institutions.

- **P.16 (*same as S.12): Provide toolkits to support implementation of partnership-focused IPE programming** — Resources will lower the barrier to starting and improving an IPE initiative because interested parties will not have to "reinvent the wheel." This, in turn, promotes buy-in. Examples include: a playbook of how IPE practitioners should connect to communities, defining best practices for building partnerships; "cheat sheets" that synthesize research findings and provide suggestions for how IPE practitioners can apply the findings in their work; a list of models or examples for how departments might incorporate IPE in their curriculum, major, or departmental activities and culture.

- **P.17: Organize forums and roundtable discussions to discuss the role of IPE in achieving broader educational goals throughout communities —** This will engage policy makers, community members, and education/academia leaders.

- **P.18 (*same as R.18): Support strategic messaging**— Equip members to leverage research findings to promote buy-in for IPE among funders, departments, and institutions. This can include curated lists of benefits and metrics of success to demonstrate the case (with evidence) that IPE efforts align with institutional and community values; templates, examples, and resources for recruiting local champions; and training on advocacy to policy makers.

- **P.19: Connect K-12 education standards to common IPE programming —** This will promote buy-in from K-12 educators, leaders, and parents, and ease implementation of IPE in school settings.

30    LEVER 2 — Engagement of interested, influential, and impacted parties

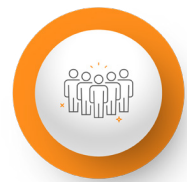

## Recommendations for (inter)national organizations

- **P.11: Acknowledge the IPE career pathways that individuals can take and legitimize the rhetoric around those IPE careers —** Science (physics) societies should actively recruit IPE professionals into their membership and can include examples of IPE careers in their career programming.

- **P.20: Physics societies should release policy statements on the importance of IPE to physics —** This will promote buy-in from members, and also provides a pathway for the organization to engage in formal advocacy on the topic.

    **Successful Example:** APS Statement on Public Engagement [67]

## Lever 3: Integrating Research-Based Practices in Informal Physics Education

One lever for achieving the desired culture change is to promote research-based IPE practices and leverage research in IPE (and informal STEM education more broadly) to understand, develop, and advocate for IPE. Integrating research-based practices can support culture change because it leads to more impactful outcomes and sustainable program design, while aligning IPE work with the scholarly tradition of physics practice. It is important to establish a clear research agenda and foster a strong research community that can help identify, disseminate, and standardize effective IPE practices, enabling their adoption across the physics community and beyond. This research can also inform program evaluation and assessment, creating a feedback loop that strengthens IPE efforts by providing insights that improve program effectiveness and demonstrate IPE's value. Increasing the body of research on informal learning serves to give the IPE community a basis for advocating for the increased centrality of IPE efforts in their departments and organizations.

Understanding IPE's role within larger systems can provide a strategic direction for specific programs. Knowing how an individual initiative fits into a broader narrative can provide clarity and purpose, guiding vision, mission, and funding strategies. Although it is not feasible to operate at a systems-level at all times, having a clear sense of focus for each program allows practitioners to make meaningful contributions within their scope. Therefore, integrating research, assessment, and evaluation provides a systematic framework for understanding audience, local community, and partners' needs, and the effectiveness and functioning of IPE programs in meeting them. This knowledge benefits practitioners, institutions, and the broader physics community.

Before presenting specific recommendations for this lever, we discuss a few general considerations that underlie all of the action items.

### 1. The role of research, assessment, and evaluation in IPE

To strengthen IPE and achieve a culture of continuous improvement, it is essential to understand the distinct roles and interconnected nature of *research*, *assessment*, and *evaluation*. Together, these components form the foundation for designing, refining, and measuring impactful programs that meet the needs of diverse audiences and partners.

While **evaluation** and **assessment** are distinct from one another, in practice these terms are often used interchangeably, and thus for the sake of this paper we combine them. The more important distinction is between research (seeking to establish new, generalizable knowledge) and evaluation/assessment (seeking to understand the impact and functioning of a particular program, allowing for iterative improvement).

Collectively, evaluation and assessment refer to the process of collecting and analyzing information to measure specific aspects of a program, particularly focused on understanding progress toward immediate outputs and goals. Evaluation/assessment can offer a comprehensive view of a program's broader impact and alignment with strategic priorities by determining the quality of present performance, asking "what" and "how" questions (i.e., What is happening? How is it happening?). In IPE, evaluation/assessment ideally occurs continuously and is often embedded within the program to provide timely insights into specific objectives. For instance, an evaluation/assessment might examine what audience a program is reaching and what immediate impacts the program has on participants. This provides valuable, real-time data to practitioners to make necessary adjustments, answering the questions, "are we accomplishing our goals?" and "how can we better meet our goals?" Evaluation and assessment are essential for justifying continued investment and scaling successful initiatives.

**Research** is an investigation aimed at generating new knowledge that contributes to a broader understanding of informal learning, public engagement, and science communication within IPE. Research in this context explores theories, frameworks, and practices to identify what drives effective informal science education and to understand the factors that foster or inhibit engagement in physics. Research is often asking "why" questions — why is the IPE program leading to certain outcomes? What does this say about informal physics education more broadly? By addressing these core questions research identifies key programmatic elements for assessment and evaluation that enable practitioners to design initiatives that are more relevant and impactful.

One practical distinction between research and evaluation/assessment is that research is typically published and requires approval from an Institutional Review Board (IRB), whereas assessment and evaluation, if conducted only for internal use, most often does not require IRB approval. Evaluation/assessment work can be published in some venues as program descriptions or case studies, in which case IRB approval or exemption may be needed.

In combination, research, assessment, and evaluation enable a more robust and sustainable approach to IPE, ensuring that programs have meaningful, long-term impact advancing the collective goals of program participants, local partners, and the broader field of physics.

## 2. Methods for research, assessment, and evaluation in IPE

Also important to the adoption of research-based practices is consideration of the methods used to research, evaluate, and assess IPE programs. Such consideration will allow the IPE community to develop protocols, instruments, tools, and techniques that provide a comprehensive understanding of the impacts of IPE participation. In order to do this, it is vital that the IPE research community develop a culture of utilizing both long-standing and novel research methods that reveal outcomes of IPE and uncover the mechanisms by which these outcomes occur. In other words, *methods matter*. Research methods matter because they constitute the all-important link between IPE models implemented (theory); the actions that occur within IPE settings (practice); and our ability to understand, improve, replicate, and scale the impacts of IPE endeavors for the benefit of all parties involved (science). The collection of methods employed in studying IPE programs dictate the knowledge generated about IPE and the extent to which this new knowledge can be generalized. IPE research tends to follow the tradition of broader physics education research in employing a variety of methods [110]. IPE research and evaluation should also draw on methods from a variety of fields including sociology, psychology, and communications. For those looking to get started in research or evaluation, an extended discussion of the variety of available research methods and data collection considerations is included in Appendix 3.

### 3. Research-Practice Partnerships (RPPs)

An excellent approach for ensuring that research is relevant and responsive to the variety of interested and impacted parties in IPE, is to engage in research-practice partnerships (RPPs). RPPs offer valuable opportunities to advance IPE by fostering collaboration between researchers and practitioners [111,112]. These partnerships are built on mutual benefit: while practitioners gain access to research that can improve their programs, researchers gain insights from real-world practice. A major benefit of this approach is that it provides a channel for iterative feedback, allowing programs to adapt and evolve based on insights from partners. By creating a continuous feedback loop, RPPs enable programs to refine their methods, maximize their impact, and respond to emerging needs in both research and practice.

However, balancing research and evaluation with authentic local community connection can be challenging. Successful RPPs require clear communication, defined goals, and an ongoing commitment to both rigor and relevance. Effective partnerships also bring a range of perspectives to the table, which can enhance both research quality and impact.

The recommendations that follow speak to how individuals, departments, informal STEM topical groups, and (inter)national organizations, respectively, can promote research-based IPE practices and leverage research in IPE to understand, develop, and advocate for IPE. The recommendations are labeled with the convention "R.#," short for "Integrating **R**esearch-based IPE Practices."

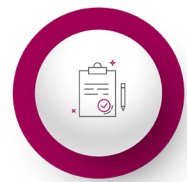

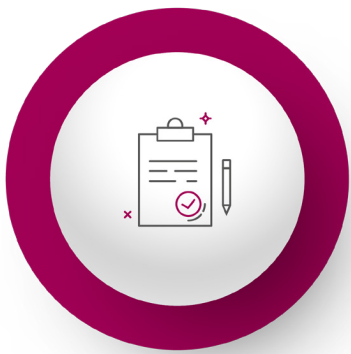

# LEVER 3
# Integrating Research-Based Practices in Informal Physics Education

Leverage research to understand, develop, and advocate for IPE as essential physics practice.

*The recommendations are labeled with the convention "R.#," short for "Integrating Research-based IPE Practices."*

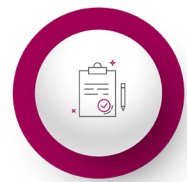

## Recommendations for individuals/individual programs

- **R.1: Clarify audience and goals —** Begin by clearly defining who you aim to serve or work with through your IPE program. Reflect on who your target audience is, what interests or needs they may have, and how your program can best meet those needs. Practitioners should consider conducting initial audience analysis to better understand what drives engagement within their communities. This could involve community feedback sessions, exploratory surveys, or informal discussions with representatives of the target audience. Identifying and articulating program goals aligned with the unique needs, backgrounds, and interests of your audience will make program design and evaluation more focused, meaningful, and effective [32].

- **R.2: Engage in continuous evaluation and assessment —** Evaluation should be an ongoing process rather than a one-time activity. Incorporate opportunities for iterative feedback from participants and partners throughout the program. This feedback can inform adjustments to the program and ensure it remains relevant and responsive to the needs of those it serves.

- **R.3: Align research with program goals, institutional missions, and community needs —** For those interested in conducting research, designing studies that align with program goals ensures that collected data will be directly relevant for assessment and evaluation. Research that explores how IPE can support both institutional missions and community needs (which should already be incorporated into program goals) offers a path forward for integrating IPE more fully into physics departments, labs, and national initiatives.

- **R.4: Seek broad expertise and partnerships —** Assemble teams with people from varied backgrounds, drawing on expertise from multiple disciplines, such as sociology, education, and communication studies. Forming research-practice partnerships (RPPs) or collaborating with other community organizations can help practitioners access a wider range of insights and resources, ensuring that programs and studies are better informed and more comprehensive.

- **R.5: Engage in dialogue rather than presenting a pitch —** Rather than presenting IPE as a "pitch" to be sold, IPE practitioners should draw on the growing body of research demonstrating its benefits to initiate dialogues that celebrate multiple perspectives and encourage collaboration — similar to the ways in which research communities typically operate. IPE provides demonstrable value to departments and organizations, and facilitators and practitioners should engage in open dialogue with the systems that could support them.

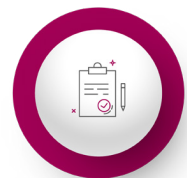

- **R.6: Define success in audience-relevant terms —** Be flexible in how success is defined, tailoring success metrics to the specific context of each program and audience. For example, if your goal is to engage younger audiences, success might focus on fostering curiosity and confidence in exploring physics. When communicating to institutions, however, metrics like participant retention and program reach might be more relevant. This approach ensures that the value of IPE is communicated effectively to different groups.

    **Example:** In an ongoing longitudinal research study, the Partnerships for Informal Science Education in the Community (PISEC) program at CU Boulder [85] tracks a range of impacts on youth participants, including sustained STEM interest and career aspirations. Researchers examine factors like STEM identity, college attendance, and persistence in STEM, while also capturing other emergent impacts reported by participants such as increased confidence in discussing science and fostering personal relationships, regardless of STEM career interest.

- **R.7: Collect a variety of metrics to tell the full story of an IPE program —** As discussed above, both quantitative and qualitative metrics are important in developing a complete picture of the impact of IPE programs for institutions, program leadership, community partners, and participants.

- **R.8: Document audience, facilitator, institutional, and community impacts —** Systematically document the impacts of IPE on the audience, the facilitators, the local community, and other impacted institutions, considering how all groups benefit from engagement. This includes tracking both quantitative metrics, like the number of participants, and qualitative indicators, such as identity development or shifts in attitudes toward science. It may not be possible or practical for one study to document the full suite of impacts, but a comprehensive picture should be built over time by the IPE community with multiple studies and evaluations. Sharing these impacts with interested, influential, and impacted parties can help make the case for the value of IPE.

- **R.9: Read IPE research digests to stay on top of latest research findings and how they apply to your practice —** Not all practitioners need to become IPE researchers, but it is important to stay abreast of how research findings can inform better practice.

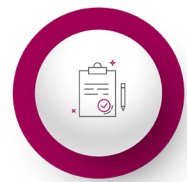

## Recommendations for departments & institutions

- **R.10: Hire faculty who research IPE —** This will indicate that the department/institution values scholarly IPE, and will facilitate the generation of new knowledge. These faculty will also train students in evaluation and research methods, expanding the research skillset of the IPE community.

- **R.11: Incorporate IPE evaluation and research into the curriculum —** Evaluation and assessment methods for both formal and informal learning environments should be included in pedagogy courses, as part of the broader body of investigation techniques.

- **R.12: Support graduate study of IPE —** Physics graduate students should be encouraged to focus on IPE research for their dissertation work, and their work with informal physics education programs should be recognized both as service and as research work.

- **R.13: Staff IRB offices with experts in the ethics and logistics of RPPs and other community-based research projects —** This will provide support for physics students and faculty who are engaging in IPE research. The IRB staff should be able to offer guidance on research involving minors and how to partner with local school districts.

- **R.14 (*same as S.7): Establish recognition mechanisms for exemplary IPE programs and practitioners, including students —** Here, exemplary refers to programs that incorporate research-based practices and evaluate the impact of their program. Such recognition could include site or practitioner awards, digital badges or professional certificates for completing IPE training, and features in institutional communications.

## Recommendations for informal STEM topical groups (e.g., JNIPER)

- **R.15 (*same as S.15): Publish IPE research "digests" —** To help the IPE community and partners stay informed of recent results and findings. This digest should be shared with science societies (e.g., APS, AAPT, AIP) and others outside the immediate IPE community.

- **R.16: Facilitate practitioner-researcher connections —** Support sustainable RPPs and foster collaborative research teams that bridge disciplinary boundaries by providing mechanisms for community members to connect and seek partnerships on evaluation and research. Connections are also needed to share methodologies, research/evaluation questions, and approaches. S.10 provides one example of how to implement this recommendation.

- **R.17: Provide training and resources around evaluation, assessment, and research —** Offer training in both qualitative and quantitative research methods, IRB support, and a repository of resources on methods, IPE literature, and theoretical frameworks. Training and resources can also include guidance on how different methods align with specific research questions, and how to handle data collection challenges. For example, if you cannot collect data from minors, you can collect retrospective data from undergraduates on their IPE experiences as minors. (This recommendation expands on the evaluation training mentioned in P. 13 (same as S.13)).

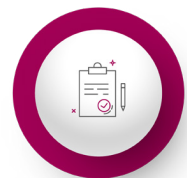

- **R.18 (*same as P.18): Support strategic messaging —** Equip members to leverage research findings in advocating for IPE with funders, departments, and institutions. This can include curated lists of research findings on benefits and metrics of success to demonstrate the case (with evidence) that IPE efforts align with institutional and community values.

- **R.19: Convene the community to define key research questions that need to be addressed —** The community should come to consensus on necessary research directions and how different research questions apply across a variety of IPE formats. This includes enumerating classical/existing questions: what are processes of change; who benefits and how; and why / how do programs work, as well as generating questions that the IPE community is just starting to ask: longitudinal impacts of IPE programs; international comparisons / context; and other novel, emergent questions.

## Recommendations for (inter)national organizations

- **R.20 (*same as S.17): Academic and commercial publishers should establish common publication venues for IPE work —** Existing venues such as *Physical Review PER* [71], *the Journal of STEM Outreach* [72], *Citizen Science: Theory & Practice* [73], *The Physics Teacher* [74], and *Connected Science Learning* [75] should better connect with the IPE community to make them aware of their publication options. Journals should also establish multiple submission types to allow for both research and programmatic articles rooted in experience, practice, and evaluation. Whenever possible, these publication venues should also be made open access.

- **R.21: Academic and commercial publishers should build infrastructure for collective data sharing —** In many cases, this is already, or soon will be, required by international and US-federal funding agencies. Publishers should require open data whenever possible, to facilitate the research community's ability to replicate and build upon prior studies.

- **R.22 (*same as S.18): Create dedicated IPE sessions at physics conferences and schedule them in prime slots -** Science societies that run physics conferences can include IPE in their abstract sorting categories, consider IPE practitioners for plenaries, and more.

- **R.23: Funding agencies should support IPE evaluation, assessment, and research —** There are at least two mechanisms for this support: (a) Provide grant lines for IPE research, as well as training for physicists in IPE research methods; (b) Require all grants to include a public engagement plan which incorporates evaluation. This evaluation should be a required component of the annual and summative grant reports, and funding withheld if evaluation is omitted. This recommendation overlaps with S.19.

## IV. Conclusion — Advancing Change in Physics through IPE

In this white paper, we have called for a culture change in physics to recognize and value IPE as an essential disciplinary practice. Informal physics education, which we also refer to as public engagement with physics, encompasses a wide variety of physics education activities beyond formal school structures. We emphasize an approach to IPE that leads to mutual engagement, authentic partnership, and promotes interactions with the public around science. In other words, the desired culture change would make physics part of local communities and local communities part of physics.

We outlined the motivations for this culture shift, putting forth that IPE makes the work of physicists relevant. It puts physics in conversation with the world we seek to better understand, inviting novel exploration and creative ideas. IPE fosters trust and supports a society where everyone benefits from science and technology advances; it serves as a gateway for entering into the physics discipline, and for staying once there; and it improves physicists' skills and research. There is precedent for this culture change, as physics has a history of redefining its disciplinary boundaries. Our call also joins similar efforts across the science community.

We then presented three levers for culture change and reported on recommendations from the 2024 JNIPER Summit regarding each of these levers. The levers include structures; engagement of interested, influential, and/or impacted parties; and integrating research-based practices in IPE. The recommendations are directed for individuals/individual programs, departments and institutions, informal STEM topical groups (e.g., JNIPER), and national and international organizations such as funders, disciplinary societies, or publishers. There is no "one-size-fits-all" prescription for enacting the desired culture change; how physicists and their institutions demonstrate a commitment to IPE as an essential physics practice will be contextually dependent. The recommendations provide a range of potential actions that can catalyze this change.

The JNIPER community has already begun to prioritize its next steps for its role in shaping the culture of physics. **We call on readers to now choose at least one recommendation directed at their level of influence (see Appendix 4) under each lever and to set forth a roadmap for implementation.** Together, with individual, institutional, and international action, the IPE community and its supporters have the power to transform physics culture, for the betterment of physics and the local communities of which we are a part.

## V. References Cited

[1] R. P. Crease and P. D. Bond, *The Leak: Politics, Activists, and Loss of Trust at Brookhaven National Laboratory* (The MIT Press, Cambridge, Massachusetts, 2022).

[2] D. Kaiser, *How the Hippies Saved Physics: Science, Counterculture, and the Quantum Revival* (W. W. Norton & Company, Erscheinungsort Nicht Ermittelbar, 2012).

[3] I. Hargittai, Leo Szilard: the physicist who envisaged nuclear weapons but later opposed their use, Phys. World (2023).

[4] E. Cartlidge, Law and the end of the world, Phys. World (2010).

[5] L. Bao, M. N. Calice, D. Brossard, B. Beets, D. A. Scheufele, and K. M. Rose, How institutional factors at US land-grant universities impact scientists' public scholarship, Public Underst. Sci. 32, 124 (2023).

[6] E. H. Ecklund, S. A. James, and A. E. Lincoln, How Academic Biologists and Physicists View Science Outreach, PLOS ONE 7, e36240 (2012).

[7] M. N. Calice, B. Beets, L. Bao, D. A. Scheufele, I. Freiling, D. Brossard, N. W. Feinstein, L. Heisler, T. Tangen, and J. Handelsman, Public engagement: Faculty lived experiences and perspectives underscore barriers and a changing culture in academia, PLOS ONE 17, e0269949 (2022).

[8] K. A. Assamagan, M. Carneiro, S. Demers, K. Jepsen, D. Lincoln, and A. Muronga, *The Need for Structural Changes to Create Impactful Public Engagement in US Particle Physics*, arXiv:2203.08916.

[9] N. C. Woitowich, G. C. Hunt, L. N. Muhammad, and J. Garbarino, Assessing motivations and barriers to science outreach within academic science research settings: A mixed-methods survey, Front. Commun. 7, (2022).

[10] J. Falk and M. Storksdieck, Using the contextual model of learning to understand visitor learning from a science center exhibition, Sci. Educ. 89, 744 (2005).

[11] J. H. Falk and L. D. Dierking, *Learning from Museums* (Rowman & Littlefield, 2018).

[12] M. B. Bennett, C. Fracchiola, D. B. Harlow, and K. Rosa, *Informal Learning in Physics*, in *The International Handbook of Physics Education Research: Teaching Physics* (AIP Publishing, Melville, New York, 2023), p. 12.1-12.28.

[13] M. Phipps, Research Trends and Findings From a Decade (1997–2007) of Research on Informal Science Education and Free-Choice Science Learning, Visit. Stud. 13, 3 (2010).

[14] L. Avraamidou and W.-M. Roth, *Intersections of Formal and Informal Science*, 1st ed. (Routledge, 2017).

[15] National Research Council, Board on Science Education, Center for Education, and Division of Behavioral and Social Sciences and Education, *Learning Science in Informal Environments: People, Places, and Pursuits* (National Academies Press, Washington, DC, 2009).

[16] C. O'Connor and J. O. Weatherall, *The Misinformation Age: How False Beliefs Spread* (Yale University Press, 2019).

[17] AAAS Center for Public Engagement with Science & Technology, *AAAS Logic Model for Public Engagement with Science*, https://www.aaas.org/sites/default/files/content_files/2016-09-15_AAAS-Logic-Model-for-Public-Engagement_Final.pdf.

[18] E. McCallie, L. Bell, T. Lohwater, J. H. Falk, J. L. Lehr, B. V. Lewenstein, C. Needham, and B. Wiehe, Many Experts, Many Audiences: Public Engagement with Science and Informal Science Education, A CAISE Inquiry Group Report, Center for Advancement of Informal Science Education (CAISE), 2009.

[19] A. V. Maltese and R. H. Tai, Eyeballs in the Fridge: Sources of early interest in science, Int. J. Sci. Educ. 32, 669 (2010).

[20] N. Oreskes, O. Edenhofer, J. Krosnick, M. S. Lindee, M. Lange, and M. Kowarsch, *Why Trust Science?* (Princeton University Press, Princeton, New Jersey, 2019).

[21] Y. Yang, T. Y. Tian, T. K. Woodruff, B. F. Jones, and B. Uzzi, Gender-diverse teams produce more novel and higher-impact scientific ideas, Proc. Natl. Acad. Sci. 119, e2200841119 (2022).

[22] A. V. Maltese and R. H. Tai, Pipeline persistence: Examining the association of educational experiences with earned degrees in STEM among U.S. students, Sci. Educ. 95, 877 (2011).

[23] K. P. Dabney, R. H. Tai, J. T. Almarode, J. L. Miller-Friedmann, G. Sonnert, P. M. Sadler, and Z. Hazari, Out-of-School Time Science Activities and Their Association with Career Interest in STEM, (2012).

[24] R. H. Tai, C. Qi Liu, A. V. Maltese, and X. Fan, *Planning Early for Careers in Science*, https://www.science.org/doi/10.1126/science.1128690.

[25] K. A. Fadigan and P. L. Hammrich, A longitudinal study of the educational and career trajectories of female participants of an urban informal science education program, J. Res. Sci. Teach. 41, 835 (2004).

[26] J. Rahm and J. C. Moore, A case study of long-term engagement and identity-in-practice: Insights into the STEM pathways of four underrepresented youths, J. Res. Sci. Teach. 53, 768 (2016).

[27] C. Thorley, Physicists and Outreach: Implications of Schools Physics Outreach Programmes from the Perspective of the Participating Physicists, Unviersity College London, 2016.

[28] B. Prefontaine, C. Mullen, J. J. Güven, C. Rispler, C. Rethman, S. D. Bergin, K. Hinko, and C. Fracchiolla, *Informal Physics Programs as Communities of Practice: How Can Program Structures Support University Students' Identities?*, arXiv:2012.03022.

[29] A. A. Bergerson, B. K. Hotchkins, and C. Furse, Outreach and Identity Development: New Perspectives on College Student Persistence, J. Coll. Stud. Retent. Res. Theory Pract. 16, 165 (2014).

[30] C. Rethman, J. Perry, J. P. Donaldson, D. Choi, and T. Erukhimova, Impact of informal physics programs on university student development: Creating a physicist, Phys. Rev.

Phys. Educ. Res. 17, 020110 (2021).

[31] J. Randolph, J. Perry, J. P. Donaldson, C. Rethman, and T. Erukhimova, Female physics students gain from facilitating informal physics programs, Phys. Rev. Phys. Educ. Res. 18, 020123 (2022).

[32] M. Shapiro, *Exploring Science Outside the Classroom Key to Inspiring Future Black Physics Majors*, https://www.teamuptogether.org/about/exploring-science-outside-the-classroom-key-to-inspiring-future-black-physics-majors.

[33] C. Fracchiolla, B. Prefontaine, M. Vasquez, and K. Hinko, *Is Participation in Public Engagement an Integral Part of Shaping Physics Students' Identity?*, in *Research and Innovation in Physics Education: Two Sides of the Same Coin*, edited by J. Guisasola and K. Zuza (Springer International Publishing, Cham, 2020), pp. 225–238.

[34] A. M. Porter, R. Y. Chu, and R. Ivie, Attrition and Persistence in Undergraduate Physics Programs, American Institute of Physics, 2024. https://ww2.aip.org/statistics/attrition-and-persistence-in-undergraduate-physics-programs

[35] American Physical Society, *APS Member Survey on Public Engagement*, https://www.aps.org/initiatives/advocate-amplify/public-engagement/public-engagement-member-survey.

[36] K. A. Hinko and N. D. Finkelstein, *Impacting University Physics Students through Participation in Informal Science*, in Vol. 1513 (2013), pp. 178–181.

[37] T. K. Carroll, E. C. Nutwell, A. D. Christy, M. B. Bennett, and N. D. Finkelstein, Facilitator Stem Teacher Identity Development Via Online Informal Stem Education During the (COVID)- 19 Era, Career Tech. Educ. Res. 48, 42 (2023).

[38] N. D. Finkelstein and L. Mayhew, *Acting in Our Own Self-Interests: Blending University and Community in Informal Science Education*, in *2008 PHYSICS EDUCATION RESEARCH CONFERENCE*, Vol. 1064 (American Institute of Physics, 2008), pp. 19–22.

[39] C. Fracchiolla, B. Prefontaine, and K. Hinko, Community of practice approach for understanding identity development within informal physics programs, Phys. Rev. Phys. Educ. Res. 16, 020115 (2020).

[40] C. Fracchiolla, N. D. Finkelstein, and K. A. Hinko, *Characterizing Models of Informal Physics Programs*, in *2018 Physics Education Research Conference Proceedings* (American Association of Physics Teachers, Washington, DC, 2019).

[41] D. Izadi, J. Willison, K. A. Hinko, and C. Fracchiolla, *Developing an Organizational Framework for Informal Physics Programs*, in *2019 Physics Education Research Conference Proceedings* (American Association of Physics Teachers, Provo, UT, 2020).

[42] D. Izadi, J. Willison, N. Finkelstein, C. Fracchiolla, and K. Hinko, Towards mapping the landscape of informal physics educational activities, Phys. Rev. Phys. Educ. Res. 18, 020145 (2022).

[43] J. Willison, D. Izadi, I. Ward, K. A. Hinko, and C. Fracchiolla, *Challenges in Study Design for Characterizing the Informal Physics Landscape*, in *2019 Physics Education

[43] *Research Conference Proceedings* (American Association of Physics Teachers, Provo, UT, 2020).

[44] American Physical Society, *JNIPER | American Physical Society*, https://www.aps.org/initiatives/advocate-amplify/public-engagement/jniper.

[45] J. Lave and E. Wenger, *Situated Learning: Legitimate Peripheral Participation* (Cambridge University Press, 1991).

[46] E. Wenger-Trayner and Wenger-Trayner, *Introduction to Communities of Practice*, https://www.wenger-trayner.com/introduction-to-communities-of-practice/.

[47] B. Trench, P. Murphy, and D. Fahy, editors , *Little Country, Big Talk: Science Communication in Ireland* (Pantaneto Press, n.d.).

[48] E. Christopherson, D. A. Scheufele, and B. Smith, The Civic Science Imperative, Stanf. Soc. Innov. Rev. 16, 46 (2018).

[49] S. D. Renoe and C. Nelson, Creating a Science-Engaged Public, Issues Sci. Technol. (2022).

[50] M. A. Kaye Elizabeth Good Christopherson, Anna Dulencin, Jon, *Science for the People*, https://nautil.us/science-for-the-people-943122/.

[51] M. N. Calice, L. Bao, B. Beets, D. Brossard, D. A. Scheufele, N. W. Feinstein, L. Heisler, T. Tangen, and Handelsman, *A Triangulated Approach for Understanding Scientists' Perceptions of Public Engagement with Science*, https://journals.sagepub.com/doi/abs/10.1177/09636625221122285.

[52] C. Fracchiolla, A. C. Lau, and N. Schrode, Fostering connection: Principles and practices for well-designed public engagement in physics, Phys. Plasmas 31, 050602 (2024).

[53] M. Cole and D. L. Consortium, *The Fifth Dimension: An After-School Program Built on Diversity* (Russell Sage Foundation, New York, 2006).

[54] *JNIPER Summit 2024 Playlist* (2024). http://www.youtube.com/playlist?list=PLgxD9DiwxLGqKZx5kcyzdRffPEHLo7dck

[55] *That's Not Physics* (2024). https://www.youtube.com/watch?v=5ITCZjb2HoQ

[56] A. P. Kaminski, *The People's Spaceship: NASA, the Shuttle Era, and Public Engagement after Apollo* (University of Pittsburgh Press, 2024).

[57] A. C. Barton, Teaching science with homeless children: Pedagogy, representation, and identity, J. Res. Sci. Teach. 35, 379 (1998).

[58] A. Tyson and B. Kennedy, Public Trust in Scientists and Views on Their Role in Policymaking, 2024. https://www.pewresearch.org/science/2024/11/14/public-trust-in-scientists-and-views-on-their-role-in-policymaking/

[59] V. Cologna et al., Trust in scientists and their role in society across 68 countries, Nat. Hum. Behav. 1 (2025).

[60] *Global Risks 2024: Disinformation Tops Global Risks 2024 as Environmental Threats Intensify*,

https://www.weforum.org/press/2024/01/global-risks-report-2024-press-release/.

[61] Research!America, *Can You Name a Living Scientist?*, https://www.researchamerica.org/sd_question/can-you-name-a-living-scientist/.

[62] National Science Board, National Science Foundation, The STEM Labor Force of Today: Scientists, Engineers, and Skilled Technical Workers, No. NSB-2021-2, 2021.

[63] Pew Research Center, Public Trust in Scientists and Views on Their Role in Policymaking, 2024.

[64] J. Koetke, K. Schumann, S. M. Bowes, and N. Vaupotič, The effect of seeing scientists as intellectually humble on trust in scientists and their research, Nat. Hum. Behav. 9, 331 (2025).

[65] Research!America, *How Important Is It for Scientists to Inform Elected Officials about Their Research and Its Impact on Society?*, https://www.researchamerica.org/sd_question/how-important-is-it-for-scientists-to-inform-elected-officials-about-their-research-and-its-impact-on-society/.

[66] Research!America, *Should Scientists Consider It Part of Their Job to Inform the Public about Their Research and Its Impact on Society?*, https://www.researchamerica.org/sd_question/should-scientists-consider-it-part-of-their-job-to-inform-the-public-about-their-research-and-its-impact-on-society/.

[67] U.S. Census Bureau, IPEDS, AIP, and APS, *Retention of Individuals Marginalized by Race/Ethnicity in Physics*, https://www.aps.org/learning-center/statistics/diversity.

[68] IPEDS and APS, *Degrees Earned by Women, by Field*, https://www.aps.org/learning-center/statistics/diversity.

[69] National Center for Education Statistics, College Enrollment Rates, U.S. Department of Education, Institute of Education Sciences, 2024.

[70] A. B. Bossmann and A. Sandner, *Women in Physics in Germany: Current Developments and Actions Taken toward Equal Opportunities*, in *AIP Conference Proceedings*, Vol. 3040 (AIP Publishing, 2023), p. 050016.

[71] C. Anteneodo, C. Brito, A. Alves-Brito, S. S. Alexandre, B. N. D'Avila, and D. P. Menezes, Brazilian physicists community diversity, equity, and inclusion: A first diagnostic, Phys. Rev. Phys. Educ. Res. 16, 010136 (2020).

[72] M. F. Parisoto, J. D. S. Mendes, É. de Mello Silva, G. Bailas, and A. C. F. dos Santos, *The Road to Gender Equity in Brazil: Small Advances and Major Setbacks*, in *AIP Conference Proceedings*, Vol. 3040 (AIP Publishing, 2023), p. 050005.

[73] S. Goswami et al., *Gender Equity in Physics in India: Interventions, Outcomes, and Roadmap*, in *AIP Conference Proceedings*, Vol. 3040 (AIP Publishing, 2023), p. 050018.

[74] C. Josten, G. Lordan, and K. Robinson, The City Quantum Summit: A Briefing on Diversity and Inclusion in the Quantum Sector, The London School of Economics and Political Science, 2023.

[75] P. Mulvey and J. Pold, Physics Bachelors: Influences and Backgrounds, American Institute

of Physics, 2024.
https://ww2.aip.org/statistics/physics-bachelors-influences-and-backgrounds

[76] M. Ong, *The Double Bind in Physics Education: Intersectionality, Equity, and Belonging for Women of Color* (Harvard Education Press, 2023).

[77] S. Hyater-Adams, N. D. Finkelstein, and K. A. Hinko, *Performing Physics: An Analysis of Design-Based Informal STEAM Education Programs*, in (2018).

[78] R. Abbasi et al., Citizen science for IceCube: Name that Neutrino, Eur. Phys. J. Plus 139, 533 (2024).

[79] L. R. House, K. Gebhardt, K. Finkelstein, E. Mentuch Cooper, D. Davis, D. J. Farrow, and D. P. Schneider, Participatory Science and Machine Learning Applied to Millions of Sources in the Hobby–Eberly Telescope Dark Energy Experiment, Astrophys. J. 975, 172 (2024).

[80] L. R. House et al., Using Dark Energy Explorers and Machine Learning to Enhance the Hobby–Eberly Telescope Dark Energy Experiment, Astrophys. J. 950, 82 (2023).

[81] L. G. Bolman and T. E. Deal, Leadership and management effectiveness: A multi-frame, multi-sector analysis, Hum. Resour. Manage. 30, 509 (1991).

[82] A. Kezar, Understanding sensemaking/sensegiving in transformational change processes from the bottom up, High. Educ. 65, 761 (2013).

[83] C. Henderson, A. L. Beach, and N. Finkelstein, *Four Categories of Change Strategies for Transforming Undergraduate Instruction*, in *Transitions and Transformations in Learning and Education*, edited by P. Tynjälä, M.-L. Stenström, and M. Saarnivaara (Springer Netherlands, Dordrecht, 2012), pp. 223–245.

[84] A. C. Lau, C. Henderson, M. Stains, M. Dancy, C. Merino, N. Apkarian, J. R. Raker, and E. Johnson, Characteristics of departments with high-use of active learning in introductory STEM courses: implications for departmental transformation, Int. J. STEM Educ. 11, 10 (2024).

[85] J. C. Corbo, D. L. Reinholz, M. H. Dancy, S. Deetz, and N. Finkelstein, Framework for transforming departmental culture to support educational innovation, Phys. Rev. Phys. Educ. Res. 12, 010113 (2016).

[86] D. L. Reinholz and N. Apkarian, Four frames for systemic change in STEM departments, Int. J. STEM Educ. 5, 3 (2018).

[87] S. Elrod and A. Kezar, *Increasing Student Success in STEM: A Guide to Systemic Institutional Change* (Association of American Colleges and Universities, 2016).

[88] S. Feola et al., STEM education institutional change projects: examining enacted approaches through the lens of the Four Categories of Change Strategies Model, Int. J. STEM Educ. 10, 67 (2023).

[89] American Association of Universities et al., Framework for Systemic Change in Undergraduate STEM Teaching and Learning, American Association of Universities, 2011.

[90] E. Price, S. Siyahhan, J. Marshall, and C. DeLeone, Mobile Making: A Research-Based

Afterschool Program Led by STEM Undergraduates Serving as Near-Peer Mentors, J. STEM Outreach 6, 1 (2023).

[91] *Mobile Making | Center for Research & Engagement in STEM Education | CSUSM*, https://www.csusm.edu/crese/research/mobilemaking.html.

[92] L. Bao, M. N. Calice, D. Brossard, D. A. Scheufele, and E. M. Markowitz, Are Productive Scientists More Willing to Engage With the Public?, Sci. Commun. 46, 65 (2024).

[93] M. Smith et al., *Informal Science Education and Career Advancement*, arXiv:2112.10623.

[94] American Physical Society, *Statement on Public Engagement*, https://www.aps.org/about/governance/statements/public-engagement.

[95] University of Texas at Austin, *General Guidelines for Promotion Review of Professional-Track Faculty*, https://utexas.app.box.com/s/8ugvlqjp60mufyq2dtwvq7ca6db94nvl.

[96] University College Dublin, *Development Framework for Faculty*, https://www.ucd.ie/hr/t4media/Development%20Framework%20for%20Faculty_Final%20310322.pdf.

[97] *The Elective Classification for Community Engagement*, https://carnegieclassifications.acenet.edu/elective-classifications/community-engagement/.

[98] *Physical Review Physics Education Research*, https://journals.aps.org/prper/.

[99] *Journal of STEM Outreach*, https://www.jstemoutreach.org/.

[100] *Citizen Science: Theory and Practice*, https://theoryandpractice.citizenscienceassociation.org.

[101] *The Physics Teacher*, https://pubs.aip.org/pte.

[102] *Connected Science Learning | NSTA*, https://www.nsta.org/connected-science-learning.

[103] Research!America Public Engagement Working Group, Recommended Strategies to Integrate Public Engagement and Civic Science Training into Graduate STEMM Education, 2024.

[104] President's Council of Advisors on Science and Technology, Letter to the President: Advancing Public Engagement with the Sciences, 2023.

[105] M. S. Reed et al., Reimagining the language of engagement in a post-stakeholder world, Sustain. Sci. 19, 1481 (2024).

[106] Indigenous Corporate Training, Inc, *9 Terms to Avoid in Communications with Indigenous Peoples*, https://www.ictinc.ca/blog/9-terms-to-avoid-in-communications-with-indigenous-peoples.

[107] K. Burn, R. Conway, A. Edwards, and E. Harries, The role of school-based research champions in a school–university partnership, Br. Educ. Res. J. 47, 616 (2021).

[108] M. T. Huber and P. Hutchings, Dynamics of Departmental Change: Lessons From a Successful STEM Teaching Initiative, Change Mag. High. Learn. 53, 41 (2021).

[109] U. K. H. Ecker, S. Lewandowsky, J. Cook, P. Schmid, L. K. Fazio, N. Brashier, P. Kendeou, E. K. Vraga, and M. A. Amazeen, The psychological drivers of misinformation

belief and its resistance to correction, Nat. Rev. Psychol. 1, 13 (2022).
[110] J. W. Creswell and J. D. Creswell, *Research Design: Qualitative, Quantitative, and Mixed Methods Approaches* (SAGE Publications, 2017).
[111] C. E. Coburn and W. R. Penuel, Research-Practice Partnerships in Education: Outcomes, Dynamics, and Open Questions, Educ. Res. 45, 48 (2016).
[112] A. L. Brown, Design Experiments: Theoretical and Methodological Challenges in Creating Complex Interventions in Classroom Settings, J. Learn. Sci. 2, 141 (1992).
[113] University of Colorado Boulder, *The Partnerships for Informal Science Education in the Community*, https://www.colorado.edu/outreach/pisec/partnerships-informal-science-education-community.
[114] John Besley and Anthony Dudo, Landscaping Overview of the North American Science Communication Training Community: Topline Takeaways from Trainer Interviews, 2017.
[115] S. Yuan, A. Dudo, and J. C. Besley, Scientific societies' support for public engagement: an interview study, Int. J. Sci. Educ. Part B 9, 140 (2019).
[116] J. Risien and R. Nilson, Landscape Overview of University Systems and People Supporting Scientists in Their Public Engagement Efforts, Oregon State University, 2018.
[117] SciPEP, Insights and Practical Considerations for Communicating Basic Science, Report for the Department of Energy Office of Science and The Kavli Foundation as part of the Science Public Engagement Partnership, OSF, 2024.
[118] Todd Newman, Christopher Volpe, Laura Lindenfeld, John Besley, Anthony Dudo, and Nicole Leavey, *Assessing Scientists' Willingness to Engage in Science Communication*, (unpublished). https://sciencecounts.org/wp-content/uploads/2019/06/Assessing-Scientist-Willingness-to-Engage-in-Science-Communication.pdf
[119] C. Volpe, E. Klein, and M. Race, Americans-Motivations-for-and-Barriers-to-Engaging-with-Science.Pdf, ScienceCounts, 2022.
[120] J. J. DeVoe, *PPPL to Lead Collaborative Center Aimed at Supporting Efforts to Bring More Underserved Communities into Plasma Science and the Fusion Energy Field*, https://www.pppl.gov/news/2023/pppl-lead-collaborative-center-aimed-supporting-efforts-bring-more-underserved.
[121] K. Young-McLear, S. Zelmanowitz, R. James, D. Brunswick, and T. DeNucci, *Beyond Buzzwords and Bystanders: A Framework for Systematically Developing a Diverse, Mission Ready, and Innovative Coast Guard Workforce*, in *2021 CoNECD Proceedings* (ASEE Conferences, Virtual - 1pm to 5pm Eastern Time Each Day, 2021), p. 36070.
[122] STEMM Opportunity Alliance, STEMM Equity and Excellence 2050 - A National Strategy for Progress and Prosperity, AAAS, 2024.
[123] L. Ding and X. Liu, *Getting Started with Quantitative Methods in Physics Education Research*, in *Getting Started in PER*, edited by Charles R. Henderson and Kathleen A.

Harper, Vol. 2 (American Association of Physics Teachers, 2012), pp. 1–42.

[124] E. Fossey, C. Harvey, F. Mcdermott, and L. Davidson, Understanding and Evaluating Qualitative Research, Aust. N. Z. J. Psychiatry 36, 717 (2002).

[125] R. Vivek and Y. Nanthagopan, Review and Comparison of Multi-Method and Mixed Method Application in Research Studies, Eur. J. Manag. Issues 29, 200 (2021).

[126] R. B. Johnson and A. J. Onwuegbuzie, Mixed Methods Research: A Research Paradigm Whose Time Has Come, Educ. Res. 33, 14 (2004).

[127] M. Brody, A. Bangert, and J. Dillon, Assessing Learning in Informal Science Contexts, National Research Council for Science Learning in Informal Environments Committee, 2008.

# Appendix 1: JNIPER Summit plenary summaries

**Brooke Smith**, Director of Science and Society at the Kavli Foundation, spoke on "Discovery Science: Communicating, Connecting, & Capacity Building." Studies funded by the Kavli Foundation have found that science communication training varies widely and is often not based in research [114]. They also report that many communication trainings fall into the deficit model (i.e., that the primary driver of scientific disengagement is lack of information, rather than attitudes, beliefs, or other social factors as they relate the individual to the scientific enterprise); that evaluation of science communication training is often shallow; and that people tend to communicate with and train people who look like them. In a study on the communication and engagement priorities of scientific societies and their members, both the societies and their members ranked listening to what others think about scientific issues last [115]. Even if scientists are trained in science communication, promotion and tenure do not reward public engagement and communication work. People who focus on the important work of science communication are in fact being pushed out of academia [116].

The Kavli Foundation and partners have also explored how communicating about basic, discovery science is different from communicating more applied work [117]. They found that the motivations/interest in science may differ for scientists and the public [118]. Overall, scientists (especially physicists) are more process- or joy-minded (i.e., science for its own sake is a Good), while the public is more payoff- or hope-minded (i.e., the Good is what science can *do*). Basic scientists lean towards the process motivation, and applied scientists tend to be more payoff-minded. However, there are other ways to communicate relevance beyond "payoff" or "usefulness." Basic scientists can communicate the relevance of their work by connecting with their audience through use of familiar examples and common values. Bringing the audience into a relationship with the process of science, instead of just the product, is important to move away from a sole focus on utility.

**Eve Klein**, Senior Advisor for Public Engagement with Science at the Association of Science and Technology Centers, gave a talk titled, "How Do Adults in the United States Think about Engaging with Science?" Klein's plenary focused on a national study investigating the motivations, interest, and barriers for adult Americans to engage with science [119]. The goal of the study was to provide science engagement practitioners insights and data to develop more effective and inclusive public engagement activities. Participants were asked what activities they were most passionate about in terms of hobbies, causes, organizations, and science interests, and inquired about the factors that motivated their participation in each. A key finding from the study was the relative importance of affective constructs — joy, imagination, and wonder — as motivators for scientific participation, alongside motivators like the development of knowledge or skills[6]. Non-white participants also articulated a stronger desire to ensure that scientific activity prevented harm — likely a function of the ways in which science as an enterprise has historically harmed these groups. Crucially, a majority of respondents expressed a desire to engage with dimensions of science beyond simply learning about the content (e.g., how modern science and ancient or traditional practice relate, and the ethical questions around new technologies).

The study asked participants to select items of interest from among a list of possible science engagement activities. The data show that many common ways to engage with science, including science centers and science-related movies and tv shows, are more likely to appeal to older and white audiences — a reminder that engagement opportunities must be developed with input from the audiences they aim to serve. Hispanic and African American respondents reported more barriers to engagement with science than white respondents, particularly in their sense of belonging and identity. In her talk, Klein reminded Summit attendees that although her data provided generalisms, it is critical not to view audiences as homogenous, and to work in partnership with local communities to determine needs, desires, and goals to most effectively engage the public.

**Shannon Greco**, Science Education Senior Program Associate at Princeton Plasma Physics Laboratory (PPPL), spoke on the development of PPPL's culture of public engagement and the ways in which its model could be adapted in her talk, "PPPL Public Engagement & Workforce Development." PPPL is a national lab funded by the U.S. Department of Energy. Crucial to Greco's message was the acknowledgement that PPPL sees its science education mission both as bidirectional and as a key element of its strategic development. Greco led Summit attendees through an overview of both the history and culture of PPPL's public engagement efforts and explained ways in which PPPL has situated itself as an engaged member of the surrounding community.

PPPL's early science education efforts might have looked, in part, similar to other institutions' initiatives — a weekend community lecture series — but even at this early stage, a key difference existed in Rush Holt, physicist, Assistant Director of the PPPL, and member of the US House of Representatives who also went on to be CEO of the American Association for the Advancement of Science. Holt championed public engagement and engaged with teachers as scholars, creating a culture of community coordination uncommon in many research labs, especially at the time. As this culture developed, PPPL built goodwill with both external community members and its own internal personnel. Programs thus acquired the necessary cultural investment to become institutionalized and became part of PPPL's *tradition* of public engagement. Extrapolated to the present day, PPPL has experienced a sizable shift in both culture and vision. The lab, for example, perceives itself as more than a place where it "just makes PhDs" to an institution focused on "developing the Fusion and Plasma Workforce," including an expanded focus on pathways that do not require PhDs and a focus on developing STEM identity in students from a wide variety of backgrounds and with a wide variety of interests. These shifts not only improve PPPL's own ability to develop partnerships, including through programs like the Plasma Network for Engagement and Training (Plasma NET, as it is more commonly known), they also align PPPL strongly with an increased focus on broader impacts and community engagement from funders such as the National Science Foundation, Department of Energy, and other federal agencies.

---

[6] This differs from what Brooke Smith presented in the first plenary – where the public was more hope/payoff minded. Eve speculated that this was due to how the question was worded— In Brooke's case, the question was about what they think about science, possibly framing respondents as outsiders. In Eve's study the question ("why do you participate") situates respondents as participants of scientific activities.

To close the talk, Greco emphasized the importance of developing a few key partnerships, with community members as fully-enfranchised co-creators. They have leveraged the connections of employees to form organic connections that can then become partnerships. Being intentional about partnerships has allowed PPPL to produce a strong, sustainable public engagement effort with tangible positive impacts to a wide variety of populations and communities, strengthening the physics enterprise. As a final example of an institutional partnership, Greco shared information about the RENEW Pathways to Fusion Collaborative Center [120], the formation of which utilized the "Healthy to Innovate" Framework [121] and could serve as a model for JNIPER organizational activity.

Overall, Greco's talk highlighted the ways in which a shifting focus from outreach to relational engagement can create a strong positive feedback loop and produce a better return on an institute's investment in broader impacts.

**Katie Hinko**, Michigan State University, spoke on "Researching informal physics education; How can we leverage research to understand, develop, and advocate for informal physics/public engagement programs." Hinko's talk focused on how we as practitioners can document the transformative aspects of public engagement, for participants and practitioners alike, in order to improve public engagement programs and advocate for support. Hinko went on to define research, assessment, and evaluation and the difference between them in terms of their purposes, audiences, and focuses. She stressed that there is no singular way to research, assess, and/or evaluate informal physics learning as they happen in such a wide variety of formats and environments.

Conducting research, assessment, and evaluation on informal learning programs is important for programs, institutions, partnerships, and funders. Studying informal learning helps programs identify the impact of their work and where there are places for improvement in their curriculum and structure. Research, assessment, and evaluation data can demonstrate the value of participating in an IPE program and this helps justify a program's existence to institutional leadership and makes the case for funding. Evaluation is also crucial to maintaining fruitful collaborations, helping to identify any areas where change is needed to better serve your partnerships. The questions asked in research, assessment, and evaluation help define the state of the field and what practitioners care about. However, there is sometimes a disconnect between researchers and practitioners. Therefore, it is important to engage in research-practice partnerships in order to build trust and understanding in informal physics education research.

Hinko also acknowledged the challenges of conducting research in informal physics environments [12,42]. For example, participation in these programs is voluntary, often short-term, and very dynamic. Programs are also often interested in their impact on science interest and identity, and this is harder to measure than traditional content knowledge. There are always special considerations when researching youth participation. Resources, time, and training for doing the evaluation and research if often lacking. However, some challenges are also opportunities, and Hinko discussed how interdisciplinary informal physics education research can be, and to keep this in mind when developing research questions and methodology.

## Appendix 2: Similar calls to action

At the time of writing, our call to action is mirrored in the following initiatives, statements, and collaborative efforts [49,103,104,122]:

In August 2023, the President's Council of Advisors on Science and Technology (PCAST) issued a letter calling for advancing public engagement with the sciences [104]. They recommend creating an office of public engagement to advise all health and science Federal agencies on advancing participatory public engagement and improving communication efforts. They also recommend that the government, "Issue a clarion call to Federal agencies to make science and technology communication and public engagement a core component of their mission and strategy." Both of these recommendations also mention the need to involve public engagement experts in enacting these changes.

Scientific societies and funders are similarly recognizing the importance of public engagement. In 2024 the STEM Opportunity Alliance, a collaboration founded by AAAS and the Doris Duke Foundation, with many supporting partners, released their national strategy for adding 20 million STEM professionals to the workforce by 2050 [122]. Informal STEM education (ISE) is mentioned in four of their five strategic pillars. For example, under the "Engagement: Nurturing Curiosity in Every Child" pillar, one of the two main goals is focused on ISE, stating that we must "Provide children and their families with equitable access to high-quality STEMM learning experiences, including in informal and technology-enabled settings" [122]. This goal reflects the importance of ISE in sparking interest in STEM education and careers. The national strategy also recognizes the role scientists play in ISE; in the "Innovation" pillar, one of their approaches calls for, "Increase[d] funding for programs and initiatives that facilitate connections and engagement between researchers and communities through citizen science, crowdsourcing, prize competitions, challenges, clinical trials and university-community research partnerships." Engaging the public with research helps ensure that advances serve the involved communities.

Research!America convened a public engagement working group, with representatives from academia and philanthropy, to create a roadmap for integrating public engagement and civic science training into graduate STEMM education [103]. Their motivations align with those described above, namely the importance of engagement for building trust and ensuring STEM serves everyone. They focus on graduate students given their interest in engaging and the training stage in their career.

A newly-adopted APS statement on public engagement targets a broad range of career stages, calling for public engagement work to be considered in hiring and advancement decisions [94]. The statement reads, "The American Physical Society (APS) commends its members who engage with the public on matters related to physics. Such activities can take many forms. APS encourages its members to pursue public engagement activities…APS urges educational institutions, national laboratories, and companies that employ physicists to recognize the high value of public engagement when making hiring, assessment, promotion, and investment decisions." This statement reflects the Society's understanding that support for public engagement work must come from both the individual and institutional levels.

All of these calls, from a wide range of organizations, reinforce the importance of Informal STEM education, and the role scientists should play in this engagement. Public engagement must be an integral practice to STEM disciplines, including physics.

# Appendix 3: Discussion of research methods for those getting started with research

## Additional details on methods for research, assessment, and evaluation in IPE

**Quantitative methods** are typically utilized to measure particular metrics, identify patterns that occur in IPE settings, or test hypotheses about IPE. They are appropriate when the goal of research, evaluation, or assessment is to examine relationships between variables, predict phenomena, or establish cause [123]. Due to the tight packaging and easily digestible nature of quantitative data, it is often favored by institutions. While quantitative research is certainly beneficial, it does not provide institutions with a complete story about IPE programming. For instance, while well-developed quantitative data can be great for predicting phenomena, they are not very good for capturing context surrounding those phenomena. Quantitative methods are most appropriate when the goal of research, assessment, or evaluation is to understand the "what" of a phenomenon occurring in IPE.

**Qualitative methods** become appropriate when the goal of IPE research, evaluation, or assessment is to understand the "why" behind a phenomenon that occurs [124]. These methods are usually employed to generate nuanced understandings, explore meaning, or understand the context surrounding phenomena in IPE settings. They are beneficial to institutions because they provide them with a snapshot of participant experiences, and personal accounts that can be used to uncover barriers, motivations, and first-hand insights regarding why particular IPE programs produce certain outcomes. Further, these methods can provide institutions with narratives and stories about the impact of IPE that cannot be captured with quantitative methods.

While qualitative and quantitative methods both have their uses, the complexities of IPE settings often necessitate that both are employed to develop a complete picture of impacts. When this is the case, it is necessary to either use **multimethod or mixed method** approaches for research, assessment, or evaluation. Multimethod approaches are best utilized when the goal of research, evaluation, or assessment is to understand issues that are multifaceted enough that they cannot be fully captured by one method. This often requires collection of qualitative and quantitative data, though those data sources need not interact with each other [125]. If the goal is to collect quantitative and qualitative data that can be used to inform each other, then it is necessary to utilize mixed methods [126]. The main goal of mixed methods data collection is to use quantitative and qualitative data in concert with each other to provide a fuller understanding of IPE impacts than either data source could by themselves. This enables data-informed decisions while keeping the human element central to IPE.

Choosing the appropriate methods for research, evaluation, or assessment of an IPE initiative requires considering the audience for the data and what metrics are important for that audience. Different data artifacts (e.g., numbers and/or narratives) address different questions and serve varying purposes depending on the audience.

## Defining Meaningful Metrics

It is crucial to consider who we are engaging with and who we aim to serve. Are we designing research (and evaluation/assessment) studies that are accessible and relevant to various publics? How can we measure what audiences care about and whether our programs resonate with them? Research that documents the impact of IPE on both participants and facilitators can provide valuable insights here. For example, key research questions might examine not only participants' understanding of physics concepts but also the experiences and personal growth of facilitators. Understanding these impacts allows IPE practitioners to develop programs that are both meaningful and sustainable. Additionally, we must acknowledge that different audiences—whether institutions, local communities, or participants—value different measures of success. Thus, the research we need to conduct depends on the local context and who we are communicating with or trying to get buy-in from. Defining success for both individual programs and IPE as a whole enables us to communicate effectively with interested, influential, and/or impacted parties, and to tailor our research for maximum relevance.

### Data collection in IPE settings

Data collection in IPE settings can be challenging due to the inherent "messiness" of IPE [127]. Recall, a defining feature of many IPE activities is learner-led activities, and this means that researchers and practitioners do not control everything that goes on in an IPE setting. They can set boundaries and structures, but the interactions are meant to be dynamic and emergent. There are also a wide variety of audiences across and sometimes within IPE programs. Understanding the needs and impact on these different audiences adds a layer of complexity to research and evaluation/assessment endeavors. Additionally, if minors are involved, special care must be taken to protect the privacy of these individuals, and to use evaluation tools that are age-appropriate. Whenever research with human subjects is conducted, it is often required for an Institutional Review Board to monitor the research plans. This requirement helps uphold the safe and ethical conduct of research, but can be a challenge to navigate for IPE practitioners who are not affiliated with an academic organization, or who come from traditional physics research backgrounds.

# Appendix 4: Recommendations organized by level of influence

# Recommendations for individuals/individual programs

## GOAL

Culture shift whereby Informal Physics Education (IPE) is widely recognized as an essential physics practice

### LEVER 1

**Structures**

Amplify structures that frame IPE as an essential activity worthy of resource allocation.

### LEVER 2

**Engagement of interested, influential, and impacted parties**

Identify techniques for recruiting and empowering people to engage in and/or support IPE as essential physics practice.

### LEVER 3

**Integrating Research-Based Practices in IPE**

Leverage research to understand, develop, and advocate for IPE as essential physics practice.

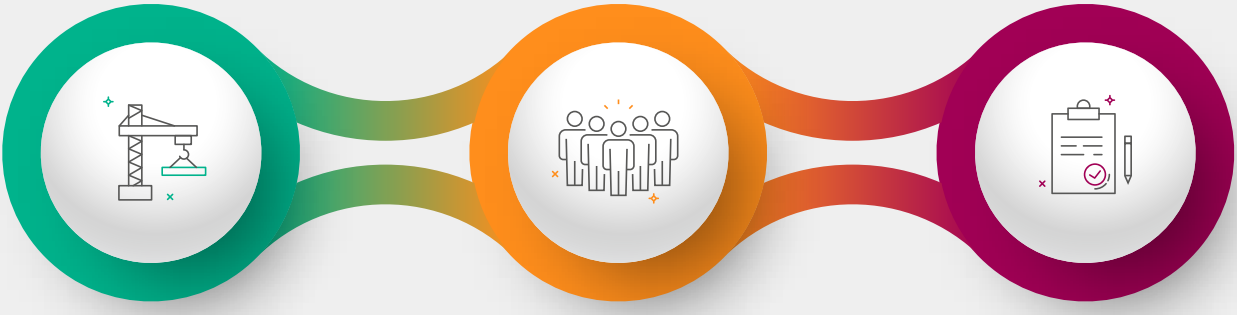

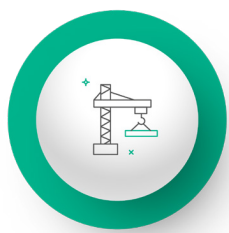

# LEVER 1 - Structures

Amplify structures that frame IPE as an essential activity worthy of resource allocation.

- **S.1:** Structural changes are often brought about by individual 'champions' who lead organizations in adopting the change. While **we encourage individuals to advocate for the following structural changes**, especially those individuals in positions of relative organizational power, **we also recognize that no one individual ought to be responsible for ensuring the following recommendations come to fruition** within their, or any, organization.

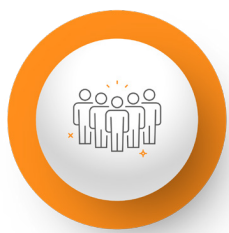

# LEVER 2 - Engagement of interested, influential, and impacted parties

Identify techniques for recruiting and empowering people to engage in and/or support IPE as essential physics practice.

- **P.1: Track program effectiveness —** University administration, funders, and others often request data on IPE program impact in order to support the effort. Sharing these impacts with interested, influential, and impacted parties can help make the case for the value of IPE. These data can be a mix of quantitative and qualitative metrics. See section III.D on research-based practices for a discussion of a variety of possible metrics.

- **P.2: Align goals with funders' missions —** Funders want to see how the project aligns with their goals. It is important to have a clear plan for the direction you want your project to go, and how support will help you get there. Additionally, proposals should incorporate rich and robust evaluation frameworks and plans for sharing evidence of impact.

- **P.3: Meet with policy makers to advocate for supportive policies and funding for IPE —** Highlight the benefits of IPE for constituencies and their connection to existing educational goals. This can be done via various mechanisms such as policy briefs, roundtables, and 1:1 meetings.

- **P.4: Invite policy makers, administrators, leaders, and faculty to interact with IPE programs —** Observing, showing up, and even participating in IPE programs will increase awareness of programs, their funding needs, and common values. These people can then be recruited as program champions, to bear witness to their importance.

- **P.5: Consistently engage community members at every stage and level of an IPE program —** Community members can be both advocates and avid participants for IPE programs. Co-creating programming with the community and your participants will lead to greater participation and impact. Co-creation can also mean leaning into the "messiness" of multiple voices and priorities.

- **P.6: Integrate IPE into your research goals —** This can take the form of a research project studying the design and impact of an IPE program, or can be incorporated into an area of traditional physics research. For example, you can engage various publics in your data collection and analysis through "community science" platforms. You can also host roundtables with community leaders to understand their needs and how they can shape your research direction.

- **P.7: When engaging in dialogue with different** interested, influential, and/or impacted parties, **articulate a relevant value proposition for supporting IPE —** This is a more general statement of a number of the other recommendations. Find the message that resonates with your audience by understanding their goals and how they connect to those of your program. For example, faculty and department leaders may be interested in the benefits to university students who facilitate IPE programs, while K-12 schools or community organizations may be interested in the benefits for younger students and their families.

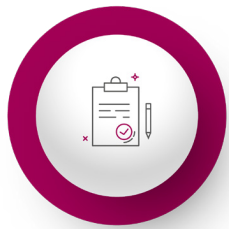

## LEVER 3 - Integrating Research-Based Practices in Informal Physics Education

Leverage research to understand, develop, and advocate for IPE as essential physics practice.

- **R.1: Clarify audience and goals —** Begin by clearly defining who you aim to serve or work with through your IPE program. Reflect on who your target audience is, what interests or needs they may have, and how your program can best meet those needs. Practitioners should consider conducting initial audience analysis to better understand what drives engagement within their communities. This could involve community feedback sessions, exploratory surveys, or informal discussions with representatives of the target audience. Identifying and articulating program goals aligned with the unique needs, backgrounds, and interests of your audience will make program design and evaluation more focused, meaningful, and effective [32].

- **R.2: Engage in continuous evaluation and assessment —** Evaluation should be an ongoing process rather than a one-time activity. Incorporate opportunities for iterative feedback from participants and partners throughout the program. This feedback can inform adjustments to the program and ensure it remains relevant and responsive to the needs of those it serves.

- **R.3: Align research with program goals, institutional missions, and community needs —** For those interested in conducting research, designing studies that align with program goals ensures that collected data will be directly relevant for assessment and evaluation. Research that explores how IPE can support both institutional missions and community needs (which should already be incorporated into program goals) offers a path forward for integrating IPE more fully into physics departments, labs, and national initiatives.

- **R.4: Seek broad expertise and partnerships —** Assemble teams with people from varied backgrounds, drawing on expertise from multiple disciplines, such as sociology, education, and communication studies. Forming research-practice partnerships (RPPs) or collaborating with other community organizations can help practitioners access a wider range of insights and resources, ensuring that programs and studies are better informed and more comprehensive.

- **R.5: Engage in dialogue rather than presenting a pitch —** Rather than presenting IPE as a "pitch" to be sold, IPE practitioners should draw on the growing body of research demonstrating its benefits to initiate dialogues that celebrate multiple perspectives and encourage collaboration — similar to the ways in which research communities typically operate. IPE provides demonstrable value to departments and organizations, and facilitators and practitioners should engage in open dialogue with the systems that could support them.

- **R.6: Define success in audience-relevant terms —** Be flexible in how success is defined, tailoring success metrics to the specific context of each program and audience. For example, if your goal is to engage younger audiences, success might focus on fostering curiosity and confidence in exploring physics. When communicating to institutions, however, metrics like participant retention and program reach might be more relevant. This approach ensures that the value of IPE is communicated effectively to different groups.

    **Example:** In an ongoing longitudinal research study, the Partnerships for Informal Science Education in the Community (PISEC) program at CU Boulder [85] tracks a range of impacts on youth participants, including sustained STEM interest and career aspirations. Researchers examine factors like STEM identity, college attendance, and persistence in STEM, while also capturing other emergent impacts reported by participants such as increased confidence in discussing science and fostering personal relationships, regardless of STEM career interest.

- **R.7: Collect a variety of metrics to tell the full story of an IPE program —** As discussed above, both quantitative and qualitative metrics are important in developing a complete picture of the impact of IPE programs for institutions, program leadership, community partners, and participants.

- **R.8: Document audience, facilitator, institutional, and community impacts —** Systematically document the impacts of IPE on the audience, the facilitators, the local community, and other impacted institutions, considering how all groups benefit from engagement. This includes tracking both quantitative metrics, like the number of participants, and qualitative indicators, such as identity development or shifts in attitudes toward science. It may not be possible or practical for one study to document the full suite of impacts, but a comprehensive picture should be built over time by the IPE community with multiple studies and evaluations. Sharing these impacts with interested, influential, and impacted parties can help make the case for the value of IPE.

- **R.9: Read IPE research digests to stay on top of latest research findings and how they apply to your practice —** Not all practitioners need to become IPE researchers, but it is important to stay abreast of how research findings can inform better practice.

# Recommendations for departments & institutions

## GOAL

Culture shift whereby Informal Physics Education (IPE) is widely recognized as an essential physics practice

### LEVER 1

**Structures**

Amplify structures that frame IPE as an essential activity worthy of resource allocation.

### LEVER 2

**Engagement of interested, influential, and impacted parties**

Identify techniques for recruiting and empowering people to engage in and/or support IPE as essential physics practice.

### LEVER 3

**Integrating Research-Based Practices in IPE**

Leverage research to understand, develop, and advocate for IPE as essential physics practice.

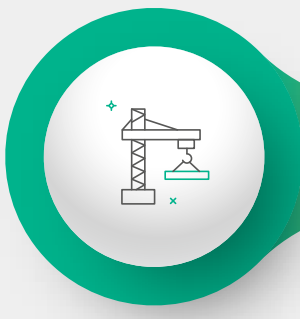
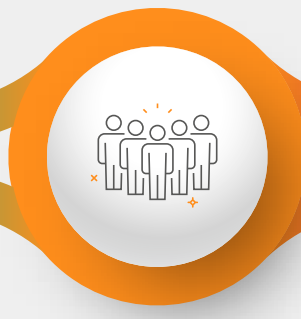
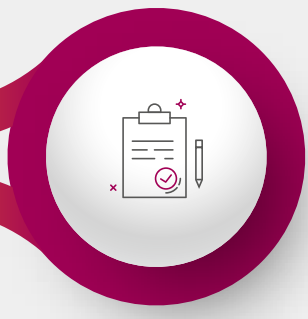

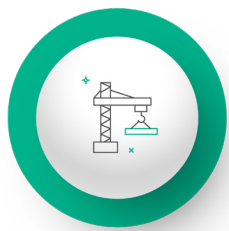

# LEVER 1 - Structures

Amplify structures that frame IPE as an essential activity worthy of resource allocation.

- **S.2: Leverage the connection between IPE and service learning** — Some undergraduate education courses are designated as "service learning" courses, in which students must engage in real-world teaching/education experience as part of the course. Physics departments and individual informal physics/STEM programs should partner with these courses to provide opportunities for students to engage in IPE and to simultaneously support the success and sustainability of the informal programs. It would be ideal if this was a structure embedded into departments and undergraduate (or even graduate) programs.

  **Successful examples:** Mobile Making at California State University [63,64] & Eric Hazlett's Analytical Physics 3 course at St Olaf College, which includes an active civic engagement component where students create hands-on demonstrations that are shared at local community events and with students in the St. Olaf TRiO Educational Talent program.

- **S.3: Integrate IPE efforts with other existing efforts to broaden participation** — Integrating IPE with existing initiatives aimed at expanding access and engagement can help departments and institutions make efficient use of established structures rather than creating new ones from scratch. This approach strengthens IPE while also fostering a more welcoming and supportive environment for a wider range of participants in physics. As with all recommendations, implementation will be highly context dependent and should attend to the organizational climate in which the department finds itself.

- **S.4 (*same as P.9): Hiring and promotion policies should include and reward IPE work** — IPE activities should not only be acknowledged in tenure and promotion cases [65], but they should be rewarded and encouraged, contrary to the existing norm in which IPE activities can hinder an individual's opportunities for tenure and promotion. For more context on this recommendation, please see the white paper published by the American Physical Society Committee on Informing the Public [66,67].

  **Successful examples:** At Lansing Community College, faculty contracts include a specified number of hours that must be dedicated on non-teaching assignments, and community outreach and events explicitly count towards this time. Dr. Bryan Stanley, LCC physics faculty, shares that in their hiring interview, they were asked about their community engagement work and plans for future community engagement they wanted to do in the position they were interviewing for.

  Similarly, the University of Texas at Austin's policy for promotion for professional track faculty (including teaching faculty) includes a statement on your primary area plus another statement on "Contributions to the Academic Enterprise" which is a broad category that can include any substantive additional work, including in IPE [68].

In Ireland, the University College Dublin 'Framework for Faculty,' used by academics in their applications for promotion, includes a specific public engagement dimension. Within this, expectations are set out for faculty with the highest levels associated with public engagement scholarship at international scale [69].

- **S.5: For faculty at academic institutions or other roles with "service" requirements, IPE activities should fulfill said service requirements —** Departments can also consider how IPE could fulfill teaching requirements if, for example, a faculty member includes an IPE component to their physics course (see recommendation S.2).

- **S.6: Provide funding for IPE work at multiple scales —** Funding need not only come from large-scale, national foundations, but should also come locally from departments, universities, etc. Funding mechanisms should prioritize IPE activities that are utilizing evidence-based practices and are designed to effectively engage the audiences of focus. Opportunities for renewable funding will help make IPE programs sustainable, and funding available to students and other junior members of institutions will help support a large population of IPE practitioners. Departments and institutions are well-suited to provide commonly needed (and relatively cheap) resources like materials, room rentals, parking waivers, and stipends for student interns.

- **S.7 (*same as R.14): Establish recognition mechanisms for exemplary IPE programs and practitioners, including students —** Such recognition could include site or practitioner awards, digital badges or professional certificates for completing IPE training, and features in institutional communications. Recognition can be implemented at the department, community of practice (e.g., JNIPER) and international organization (e.g., APS) scale.

- **S.8: Create or expand community engagement/campus extension offices —** These offices are often staffed by experts in community partnerships whose job it is to build relationships with local community members and groups. These offices can be a resource for physics faculty and students seeking to engage in IPE, as well as local audiences interested in STEM. This recommendation relates to institutions ensuring these offices have the capacity and expertise to support these kinds of connections that would be beneficial for physics/STEM public engagement.

- **S.9 (*same as P.12): Obtain the Community Engagement Elective Carnegie Classification for your institution [70]—** This classification requires a campus-wide commitment to partnership with the local community. Applications require detailed examples of academic-community partnerships, such as an IPE program. A push from institutional leaders to obtain this classification for the institution will promote buy-in from multiple interested, influential, and/or impacted parties in IPE.

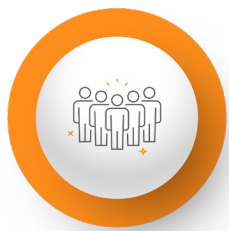

# LEVER 2 - Engagement of interested, influential, and impacted parties

Identify techniques for recruiting and empowering people to engage in and/or support IPE as essential physics practice.

- **P.8: Invest in infrastructure that supports public engagement and evaluation —** Reducing administrative and resource barriers facilitates buy-in from faculty, staff, and students. Examples of infrastructure include dedicated space and materials that can be used for IPE programming, partnerships with social scientists for program evaluation, and teaching assistant lines dedicated for facilitating IPE programs.

- **P.9 (*same as S.4): Recognize IPE activities in tenure and promotion processes —** This is important for faculty and staff to buy-in to IPE as a practice they should engage in [65–67].

- **P.10: Integrate IPE activities into the curriculum and provide academic credit for participation in IPE efforts —** This provides an important marker of disciplinary value to students. It is easier for students to buy-in to IPE as an essential disciplinary practice if they get academic credit for their participation. IPE can be incorporated into curricula as part of a service-learning credit, a lab course, a pedagogical/communication requirement, or an independent study.

    **Successful example:** Eric Hazlett's Analytical Physics 3 course at St Olaf College, which includes an active civic engagement component where students create hands-on demonstrations that are shared at local community events and with students in the St. Olaf TRiO Educational Talent program.

- **P.3: Meet with policy makers to advocate for supportive policies and funding for IPE —** Highlight the benefits of IPE for constituencies and their connection to existing educational goals. This can be done via various mechanisms such as policy briefs, roundtables, and 1:1 meetings.

- **P.4: Invite policy makers, administrators, leaders, and faculty to interact with IPE programs —** Observing, showing up, and even participating in IPE programs will increase awareness of programs, their funding needs, and common values. These people can then be recruited as program champions, to bear witness to their importance.

- **P.11: Acknowledge the IPE career pathways that students can take and legitimize the rhetoric around those IPE careers —** Career panels and talks should include examples of physics majors who have gone on to a career in IPE, including the option to pursue the academic track with a research-focus on IPE. Job boards and career resources should also illustrate the many forms an IPE career can take.

- **P.12 (*same as S.9): Obtain the Community Engagement Elective Carnegie Classification for your institution [70] —** This classification requires a campus-wide commitment to partnership with the local community. Applications require detailed examples of academic-community partnerships, such as an IPE program. A push from institutional leaders to obtain this classification for the institution will promote buy-in from multiple interested, influential, and/or impacted parties in IPE.

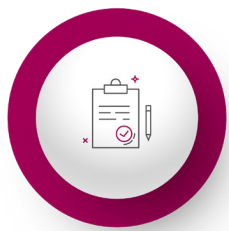

# LEVER 3 - Integrating Research-Based Practices in Informal Physics Education

Leverage research to understand, develop, and advocate for IPE as essential physics practice.

- **R.10: Hire faculty who research IPE —** This will indicate that the department/institution values scholarly IPE, and will facilitate the generation of new knowledge. These faculty will also train students in evaluation and research methods, expanding the research skillset of the IPE community.

- **R.11: Incorporate IPE evaluation and research into the curriculum —** Evaluation and assessment methods for both formal and informal learning environments should be included in pedagogy courses, as part of the broader body of investigation techniques.

- **R.12: Support graduate study of IPE —** Physics graduate students should be encouraged to focus on IPE research for their dissertation work, and their work with informal physics education programs should be recognized both as service and as research work.

- **R.13: Staff IRB offices with experts in the ethics and logistics of RPPs and other community-based research projects —** This will provide support for physics students and faculty who are engaging in IPE research. The IRB staff should be able to offer guidance on research involving minors and how to partner with local school districts.

- **R.14 (*same as S.7): Establish recognition mechanisms for exemplary IPE programs and practitioners, including students —** Here, exemplary refers to programs that incorporate research-based practices and evaluate the impact of their program. Such recognition could include site or practitioner awards, digital badges or professional certificates for completing IPE training, and features in institutional communications.

# Recommendations for informal STEM topical groups (e.g., JNIPER)

## GOAL

Culture shift whereby Informal Physics Education (IPE) is widely recognized as an essential physics practice

### LEVER 1

**Structures**

Amplify structures that frame IPE as an essential activity worthy of resource allocation.

### LEVER 2

**Engagement of interested, influential, and impacted parties**

Identify techniques for recruiting and empowering people to engage in and/or support IPE as essential physics practice.

### LEVER 3

**Integrating Research-Based Practices in IPE**

Leverage research to understand, develop, and advocate for IPE as essential physics practice.

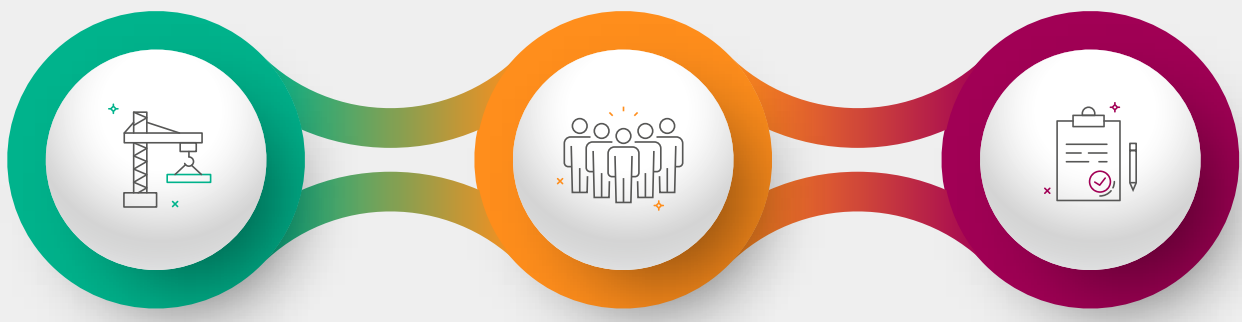

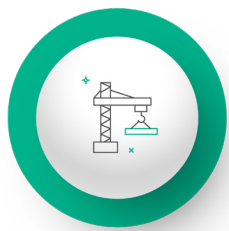

# LEVER 1 - Structures

Amplify structures that frame IPE as an essential activity worthy of resource allocation.

- **S.10: Host a member directory or other mechanism for facilitating partnerships among IPE practitioners, IPE researchers, and other physicists**

- **S.11: Publish a regular newsletter to support community building, help disseminate members' work, and provide visibility and recognition**

- **S.12 (\*same as P.16): Provide toolkits and training to support implementation of partnership-focused IPE programming —** Resources will lower the barrier to starting and improving an IPE initiative because interested parties will not have to "reinvent the wheel." Examples include: a playbook of how IPE practitioners should connect to communities, defining best practices for building partnerships; "cheat sheets" that synthesize research findings and provide suggestions for how IPE practitioners can apply the findings in their work; a list of models or examples for how departments might incorporate IPE in their curriculum, major, or departmental activities and culture.

- **S.13 (\*same as P.13): Provide workshops on effective grant writing, impact reporting, and evaluation strategies —** This training can help individuals prepare an IPE-focused grant proposal, and/or help individuals to directly connect their physics research meaningfully to their broader impacts components of their physics research grants.

- **S.14: Set a recognized standard for ethical and effective IPE practice and research —** These standards can be emulated and boosted by science societies, funders, and other IPE partners. Examples of standards include paying and/or recognizing students facilitating IPE programs; cultivating long-term community partnerships; engaging in research-practice partnerships; and establishing clear metrics and evaluation plans. For more on standards for research, see Section III.D.

- **S.7 (\*same as R.14): Establish recognition mechanisms for exemplary IPE programs and practitioners, including students —** Such recognition could include site or practitioner awards, digital badges or professional certificates for completing IPE training, and features in institutional communications. Recognition can be implemented at the department, community of practice (e.g., JNIPER) and international organization (e.g., APS) scale.

- **S.15 (\*same as R.15): Publish IPE research "digests" —** To help the IPE community and partners stay informed of recent results and findings. This digest should be shared with science societies (e.g., APS, AAPT, AIP) and others outside the immediate IPE community.

- **S.16: Create and disseminate mentorship opportunities —** To mentor people new to IPE into career pathways that incorporate IPE.

   **Successful example:** The JNIPER Fellows program trains a small cohort of students in science communication skills [24]. The students then apply their skills to produce content for APS Public Engagement programs. The cohort design and connection to the broader JNIPER community provide exposure to multiple pathways that incorporate IPE.

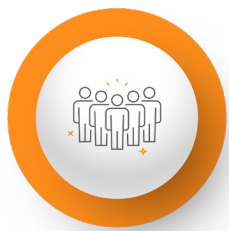

# LEVER 2 - Engagement of interested, influential, and impacted parties

Identify techniques for recruiting and empowering people to engage in and/or support IPE as essential physics practice.

- **P.13 (*same as S.13):** Provide workshops on effective grant writing, impact reporting, and evaluation strategies — This training can help individuals prepare an IPE-focused grant proposal, and/or help individuals to directly connect their physics research meaningfully to their broader impacts components of their physics research grants. Training in evaluation helps practitioners know what to assess (e.g., participant value, motivations) and helps with messaging to funders and leaders because they often require metrics of IPE program impact.

- **P.14: Highlight successful IPE programs** — Concrete examples of programs that are meeting their goals and those of their community partners help promote buy-in from parties who have not experienced an IPE initiative. Similarly, sharing stories of challenges and how programs have overcome them will add to buy-in.

- **P.15: Highlight institutional practices that uplift IPE work** — Collect concrete examples of institutions that are recognizing the IPE work of their students/staff/faculty and examples how they are supporting the work. This helps facilitate change at similar type institutions.

- **P.16 (*same as S.12):** Provide toolkits to support implementation of partnership-focused IPE programming — Resources will lower the barrier to starting and improving an IPE initiative because interested parties will not have to "reinvent the wheel." This, in turn, promotes buy-in. Examples include: a playbook of how IPE practitioners should connect to communities, defining best practices for building partnerships; "cheat sheets" that synthesize research findings and provide suggestions for how IPE practitioners can apply the findings in their work; a list of models or examples for how departments might incorporate IPE in their curriculum, major, or departmental activities and culture.

- **P.17: Organize forums and roundtable discussions to discuss the role of IPE in achieving broader educational goals throughout communities** — This will engage policy makers, community members, and education/academia leaders.

- **P.18 (*same as R.18):** Support strategic messaging— Equip members to leverage research findings to promote buy-in for IPE among funders, departments, and institutions. This can include curated lists of benefits and metrics of success to demonstrate the case (with evidence) that IPE efforts align with institutional and community values; templates, examples, and resources for recruiting local champions; and training on advocacy to policy makers.

- **P.19: Connect K-12 education standards to common IPE programming** — This will promote buy-in from K-12 educators, leaders, and parents, and ease implementation of IPE in school settings.

69                                       APPENDIX — Recommendations for informal STEM topical groups (e.g., JNIPER)

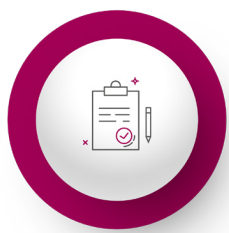

# LEVER 3 - Integrating Research-Based Practices in Informal Physics Education

Leverage research to understand, develop, and advocate for IPE as essential physics practice.

- **R.15 (*same as S.15): Publish IPE research "digests"** — To help the IPE community and partners stay informed of recent results and findings. This digest should be shared with science societies (e.g., APS, AAPT, AIP) and others outside the immediate IPE community.

- **R.16: Facilitate practitioner-researcher connections** — Support sustainable RPPs and foster collaborative research teams that bridge disciplinary boundaries by providing mechanisms for community members to connect and seek partnerships on evaluation and research. Connections are also needed to share methodologies, research/evaluation questions, and approaches. S.10 provides one example of how to implement this recommendation.

- **R.17: Provide training and resources around evaluation, assessment, and research** — Offer training in both qualitative and quantitative research methods, IRB support, and a repository of resources on methods, IPE literature, and theoretical frameworks. Training and resources can also include guidance on how different methods align with specific research questions, and how to handle data collection challenges. For example, if you cannot collect data from minors, you can collect retrospective data from undergraduates on their IPE experiences as minors. (This recommendation expands on the evaluation training mentioned in P. 13 (same as S.13)).

- **R.18 (*same as P.18): Support strategic messaging** — Equip members to leverage research findings in advocating for IPE with funders, departments, and institutions. This can include curated lists of research findings on benefits and metrics of success to demonstrate the case (with evidence) that IPE efforts align with institutional and community values.

- **R.19: Convene the community to define key research questions that need to be addressed** — The community should come to consensus on necessary research directions and how different research questions apply across a variety of IPE formats. This includes enumerating classical/existing questions: what are processes of change; who benefits and how; and why / how do programs work, as well as generating questions that the IPE community is just starting to ask: longitudinal impacts of IPE programs; international comparisons / context; and other novel, emergent questions.

# Recommendations for (inter)national organizations

## GOAL

Culture shift whereby Informal Physics Education (IPE) is widely recognized as an essential physics practice

### LEVER 1

**Structures**

Amplify structures that frame IPE as an essential activity worthy of resource allocation.

### LEVER 2

**Engagement of interested, influential, and impacted parties**

Identify techniques for recruiting and empowering people to engage in and/or support IPE as essential physics practice.

### LEVER 3

**Integrating Research-Based Practices in IPE**

Leverage research to understand, develop, and advocate for IPE as essential physics practice.

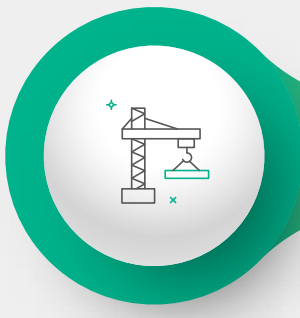
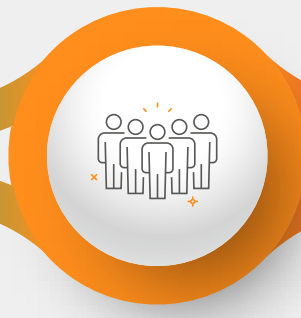
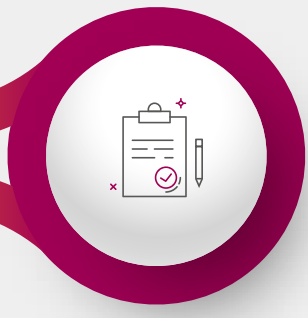

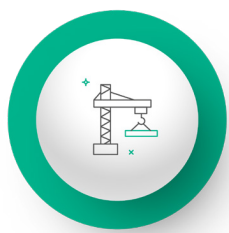

# LEVER 1 - Structures

Amplify structures that frame IPE as an essential activity worthy of resource allocation.

- **S.17 (\*same as R.20): Academic and commercial publishers should establish common publication venues for IPE work —** Existing venues such as *Physical Review PER* [71], the *Journal of STEM Outreach* [72], *Citizen Science: Theory & Practice* [73], *The Physics Teacher* [74], and *Connected Science Learning* [75] should better connect with the IPE community to make them aware of their publication options. Journals should also establish multiple submission types to allow for both research and programmatic articles rooted in experience, practice, and evaluation. Whenever possible, these publication venues should also be made open access.

- **S.7 (\*same as R.14):** Establish recognition mechanisms for exemplary IPE programs and practitioners, including students — Such recognition could include site or practitioner awards, digital badges or professional certificates for completing IPE training, and features in institutional communications. Recognition can be implemented at the department, community of practice (e.g., JNIPER) or international organization (e.g., APS) scale.

- **S.18 (\*same as R.22):** Create dedicated IPE sessions at physics conferences and schedule them in prime slots — Science societies that run physics conferences can include IPE in their abstract sorting categories, consider IPE practitioners for plenaries, and more.

- **S.19: Funding agencies should support IPE through the following mechanisms —** Require all grants to include a public engagement plan, and providing detailed guidance on evidence-based best practices; Attend to public engagement action and implementation in annual grant reports and reviews, ensuring that quality IPE is valued and is not relegated to a "box-checking" exercise; Fund IPE programs directly; Fund studies of public engagement and science communication trainings to expand knowledge of effective training practices; Direct graduate fellowship awards to include funding for public engagement training and implementation. (This last example is aligned with the call from the Research!America Public Engagement Working Group [76] .)  This recommendation overlaps with R.23.

- **S.20: Create federal structures that support national agencies in growing their participatory public engagement and science communication efforts—** See, for example, the August 2023 Letter from the Presidential Council of Advisors on Science and Technology [77].

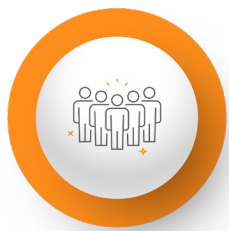

## LEVER 2 - Engagement of interested, influential, and impacted parties

Identify techniques for recruiting and empowering people to engage in and/or support IPE as essential physics practice.

- **P.11: Acknowledge the IPE career pathways that students can take and legitimize the rhetoric around those IPE careers —** Career panels and talks should include examples of physics majors who have gone on to a career in IPE, including the option to pursue the academic track with a research-focus on IPE. Job boards and career resources should also illustrate the many forms an IPE career can take.

- **P.20: Physics societies should release policy statements on the importance of IPE to physics —** This will promote buy-in from members, and also provides a pathway for the organization to engage in formal advocacy on the topic.

    **Successful Example:** APS Statement on Public Engagement [67].

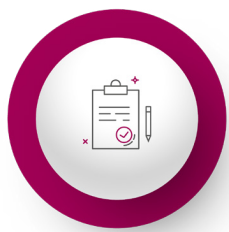

## LEVER 3 - Integrating Research-Based Practices in Informal Physics Education

Leverage research to understand, develop, and advocate for IPE as essential physics practice.

- **R.20 (*same as S.17):** Academic and commercial publishers should establish common publication venues for IPE work — Existing venues such as *Physical Review PER* [71], the *Journal of STEM Outreach* [72], *Citizen Science: Theory & Practice* [73], *The Physics Teacher* [74], and *Connected Science Learning* [75] should better connect with the IPE community to make them aware of their publication options. Journals should also establish multiple submission types to allow for both research and programmatic articles rooted in experience, practice, and evaluation. Whenever possible, these publication venues should also be made open access.

- **R.21: Academic and commercial publishers should build infrastructure for collective data sharing —** In many cases, this is already, or soon will be, required by international and US-federal funding agencies. Publishers should require open data whenever possible, to facilitate the research community's ability to replicate and build upon prior studies.

- **R.22 (*same as S.18): Create dedicated IPE sessions at physics conferences and schedule them in prime slots -** Science societies that run physics conferences can include IPE in their abstract sorting categories, consider IPE practitioners for plenaries, and more.

- **R.23: Funding agencies should support IPE evaluation, assessment, and research —** There are at least two mechanisms for this support: (a) Provide grant lines for IPE research, as well as training for physicists in IPE research methods; (b) Require all grants to include a public engagement plan which incorporates evaluation. This evaluation should be a required component of the annual and summative grant reports, and funding withheld if evaluation is omitted. This recommendation overlaps with S.19.